\documentclass[fleqn,usenatbib]{mnras}

\usepackage{newtxtext,newtxmath}
\usepackage[T1]{fontenc}

\DeclareRobustCommand{\VAN}[3]{#2}
\let\VANthebibliography\thebibliography
\def\thebibliography{\DeclareRobustCommand{\VAN}[3]{##3}\VANthebibliography}

\usepackage{graphicx,xcolor,ulem}	% Including figure files
\usepackage{amsmath}	% Advanced maths commands
\usepackage{breqn}
\title[WInDI]{WInDI: a Warp-Induced Dust Instability in protoplanetary discs}

\author[Aly et al.]{
Hossam Aly,$^{1,2}$\thanks{E-mail: hossam.saed@gmail.com}
Rebecca Nealon,$^{3,4}$
and Jean-François Gonzalez$^{5}$
\\
$^{1}$Faculty of Aerospace Engineering, Delft University of Technology, Kluyverweg 1, 2629 HS Delft, The Netherlands\\
$^{2}$Zentrum für Astronomie der Universität Heidelberg, Astronomisches Rechen-Institut, Mönchhofstr. 12-14, 69120 Heidelberg, Germany \\
$^{3}$Centre for Exoplanets and Habitability, University of Warwick, Coventry CV4 7AL, UK\\
$^{4}$Department of Physics, University of Warwick, Coventry CV4 7AL, UK \\
$^{5}$Universite Claude Bernard Lyon 1, CRAL UMR5574, ENS de Lyon, CNRS, Villeurbanne, F-69622, France\\
}

\date{Accepted XXX. Received YYY; in original form ZZZ}

\pubyear{2015}

\begin{document}
\label{firstpage}
\pagerange{\pageref{firstpage}--\pageref{lastpage}}
\maketitle

\begin{abstract}
We identify a new dust instability that occurs in warped discs. The instability is caused by the oscillatory gas motions induced by the warp in the bending wave regime. We first demonstrate the instability using a local 1D (vertical) toy model based on the warped shearing box coordinates and investigate the effects of the warp magnitude and dust Stokes number on the growth of the instability. We then run 3D SPH simulations and show that the instability is manifested globally by producing unique dust structures that do not correspond to gas pressure maxima. The 1D and SPH analysis suggest that the instability grows on dynamical timescales and hence is potentially significant for planet formation.
\end{abstract}


%%%%%%%%%%%%%%%%%%%%%%%%%%%%%%%%%%%%%%%%%%%%%%%%%%

%%%%%%%%%%%%%%%%% BODY OF PAPER %%%%%%%%%%%%%%%%%%

\section{Introduction}

Warped or misaligned discs (those with a curvature that varies as a function of radius) are most often observed in scattered light where the inner warped structure is inferred from the shadow it casts on the outer disc \citep[e.g.][]{Casassus:2015yu}. Strong misalignments (with inclinations of $\gtrsim 45 \deg$) have been observed in scattered light observations since HD~142527 in numerous discs \citep[e.g.][]{Casassus:2015yu,Benisty:2017kq,Benisty:2018ve,Kraus:2020nw}. Smaller misalignments ($\lesssim 10 \deg$) have also been identified \citep{Debes:2017fk,Muro-Arena:2020bs} however \citet{Young:2022vh} caution that these smaller warps are difficult to identify and thus may have been missed in existing data. Some of these disc warps may be caused by interactions with external misaligned companions \citep{Gonzalez:2020bh,Nealon:2020vg}, by the misaligned orbits of the stars interior to the disc \citep{Benisty:2018ve,Kraus:2020nw,Smallwood:2021nm} or by subsequent accretion episodes with different angular momentum to the existing disc \citep{Bate:2018ls}. With the growing weight of observational evidence of warps in protoplanetary discs it is clear that warps are likely common.

How the warp is communicated through the disc depends on comparison of the \citet{shakura_sunyaev} $\alpha$ viscosity to the disc aspect ratio, given as the ratio of the scale height and radius $H/R$. In discs surrounding black holes typically $\alpha \gtrsim H/R$ and the relatively high viscosity of the disc moderates the internal motions caused by the oscillating pressure gradient, causing the warp to undergo a diffusive, rather than wavelike, evolution \citep{pap_pringle_1983}. The alternative is expected for protoplanetary discs with $\alpha \lesssim H/R$, causing the warp to propagate as a bending wave with half the local sound speed, $c_s(R)/2$ and very little damping of the warp occurs \citep{Papaloizou:1995fk}. \citet{pringle_1992} used the conservation of mass and angular momentum to describe the \textit{global} evolution of a warped disc. In doing so they utilised two important concepts. First, that we can conceptually understand warps by envisaging the disc to be constructed of concentric rings \citep[see][for a schematic of this]{Ogilvie:2013by}, with the properties of each ring (e.g. surface density, $\Sigma (R,t)$ and unit angular momentum vector $\ell (R,t)$) being a function of its distance from the central object, $R$ and time $t$ \citep[but see also][]{pap_pringle_1983}. Second, that the disc viscosity $\nu$ becomes more complicated in the presence of the warp with multiple components; $\nu_1$, $\nu_2$ and later introduced by \citet{Ogilvie:1999lr} $\nu_3$. Importantly, the $\nu_2$ and $\nu_3$ coefficients are determined by the internal oscillatory shear flows that are responsible for the transport of angular momentum \citep{pap_pringle_1983}. Analytic approaches make use of this framework to consider the evolution of warped discs in the presence of black holes \citep{lubow2002evolution}, binaries \citep{facchini_2013} and interpretations of disc breaking \citep{Dogan:2018wo}. Until recently, the analytic prescriptions for the wave-like and diffusive regimes were treated completely separately. \citet{Martin:2019hj} generalised both sets of equations such that the internal torque $G$ (responsible for the viscous transport of mass) vanished in the correct regime. They also added a $\beta$ term that damped the spurious evolution of the surface density from their generalised equations. Beginning from the shearing box prescription of \citet{Ogilvie:2013by}, \citet{Dullemond:2022ng} derived a set of equations valid in both regimes but recovering the spurious evolution of the surface density profile. They then fixed the internal torque vector $G$ to co-rotate with the orientation vector, overcoming this issue.

Less frequently observed than warps, young planets have also been directly found in observations of protoplanetary discs e.g. PDS70b and c \citep[][]{Benisty:2021bh} and most recently HD~169142b \citep{Hammond:2023gu}. Commonly observed ring and gap dust substructure \citep[e.g.][]{Andrews:2016bw} have also been proposed as signposts of planet formation, particularly as subsequent work by \citet{Pinte:2018ah,Pinte:2020gh} has shown that dust gaps are often coincident with kinematic deviations caused by a planet away from the assumed Keplerian profile. Importantly, these observations show evidence of planet-disc interactions rather than planet formation and those that do show planet formation (e.g. PDS70) have proven to be quite rare. In combination with implied short disc life-times, this suggests that planet formation is likely to either be rapid or to begin earlier than previously expected.

The onset of planet formation in a dust rich protoplanetary disc may be motivated through instabilities in the dust. \citet{Squire:2018bb} identified the family of resonant drag instabilities (RDIs), where an instability occurs whenever the frequency of the fluid wave matches the pattern speed of the dust drifting along that wave. The streaming instability forms one family of these RDIs, where the gas and dust interact via aerodynamic drag causing radial drift of the dust and initialises instabilities \citep{Youdin:2005by}. The settling instability is the most general application of the streaming instability, incorporating both the vertical settling of dust grains towards the mid-plane and radial drift. In the case of a viscosity around $\alpha \sim 10^{-6}$ this results in dust concentrations that are suggestive for planet formation and for this process to be efficient, it additionally requires $\gtrsim 1$ dust-to-gas mass ratios \citep{Youdin:2005by}. However, the dust settling instability is limited to these conditions and additionally performs poorly for quite small grains \citep{Drazkowska:2014vq} and tends to be weak in 3D simulations \citep{Squire:2018bb}. Indeed, \citet{Krapp:2020cq} suggest that the settling instability is not a favoured pathway for dust collection necessary for planet formation.

In addition to the above dust instabilities, warped discs may be vulnerable to the parametric instability. \citet{Papaloizou:1995fk} initially suggested that the parametric resonance of inertial waves would make the oscillatory horizontal flow in such a disc unstable. \citet{Ogilvie:2013lw} and \citet{Paardekooper:2019ns} used a warped local shearing box model to demonstrate exactly this, finding that the parametric instability was comprehensively experienced by a warped disc leading to rolling flows and the formation of rings in the gas. However, the parametric instability is difficult to recover in global disc simulations due to the dual requirements of high numerical resolution and low $\alpha$ viscosity \citep{Ogilvie:2013lw}. Indeed, these stringent conditions are not met by the simulations we present in this work. Dedicated efforts have recently recovered the parametric instability in a global disc by using up to 120 million particles \citep{Deng:2021fq} and in local simulations \citep{Fairburn:2023as}.

In this paper we unite the ideas of warps, dust and instabilities to consider how a warp changes the dust dynamics in protoplanetary discs. Here, we build on previous work looking at gas and dust dynamics in non-coplanar discs that have identified dust `traffic jams' \citep{Aly:2020ds,Aly:2021bu,Longarini:2021ns} but restrict ourselves to scenarios where disc breaking is not expected. In answering this question we uncover the new Warp Induced Dust Instability (WInDI) which we consider from both a theoretical and numerical perspective. As we shall show, this instability occurs in the presence of a warp due to the drag between the dust and gas. In Section~\ref{section:recap} we summarise the warped shearing box framework that governs the local dynamics of warped discs and extend it to include dust. In Section~\ref{section:1D_results} we motivate this instability using a 1D toy model based on the warped shearing box framework. In Section~\ref{section:3D_results} we recover WInDI in global 3D smoothed particle hydrodynamic simulations. In Section~\ref{section:discussion} we contextualise WInDI with respect to other instabilities, its relevance to planet formation, and present some of the caveats of this study as well as outline avenues for future extensions. We present our conclusions in Section~\ref{section:conclusion}.

\section{Analytic framework}
\label{section:recap}
The local hydrodynamics of accretion discs are captured by the standard shearing box treatment \citep[e.g.][]{hawley_gammie_balbus_1995}. In this local treatment, a co-moving small patch of gas is described as a Cartesian coordinate system with $x$, $y$ and $z$ corresponding to the radial, azimuthal and vertical directions. Assuming an isothermal equation of state, the gas pressure $p$ is related to the density $\rho$ by $p = c_s^2 \rho$, where $c_s$ is the constant isothermal sound speed. Neglecting self-gravity and the effects of magnetic fields, gas with velocity $\Vec{u}$ is governed by the hydrodynamical equations:
\begin{align}
    \mathrm{D}u_x - 2\Omega_0 u_y &= 2q\Omega_0^2 x - (1/\rho)\partial_x p,\\
    \mathrm{D}u_y + 2\Omega_0 u_x &= - (1/\rho) \partial_y p,\\
    \mathrm{D}u_z &= - \Omega_0^2 z - (1/\rho) \partial_z p,\\
    \mathrm{D}\rho &= - \rho ( \partial_x u_x + \partial_y u_y + \partial_z u_z),
\end{align}
where $\Omega_0$ is the angular velocity with a reference radius $r_0$. $q$ is the orbital shear rate $q=-\mathrm{d} \ln{\Omega}/\mathrm{d} \ln{r}$ ($q=1.5$ for a Keplerian orbit), and D is the Lagrangian derivative
\begin{align}
    \mathrm{D} = \partial_t + u_x \partial_x + u_y \partial_y + u_z \partial_z .
\end{align}
 We note that the first term on the right hand side of the $x$ momentum equation (i.e., radial) is the centrifugal force and the second term on the left hand side of the $x$ and $y$ momentum equations represent the Coriolis force. The first term on the
right hand side of the vertical momentum equation is the vertical component of the gravitational force, which in case of hydrostatic equilibrium is balanced by the vertical pressure gradient resulting in a Gaussian vertical density profile. We now consider how this framework can be expanded to include both dust and the effect of a global disc warp.

\subsection{Warped shearing box}
\citet{Ogilvie:2013by} introduce the warped shearing coordinates which follow the orbital motion of the shearing box as it traverses an orbit in the warped disc (see their Figure 1 for an illustrative schematic). These are related to the unwarped Cartesian coordinates via the transformations
\begin{align}
    t' &= t,\\
    x' &= x,\\
    y' &= y + q\tau x,\\
    z' &= z + |\psi|\cos({\tau})x,
\end{align}
 where $\tau = \Omega_0 t'=\Omega_0 t$ is the orbital phase, and $\psi$ is the dimensionless warp amplitude defined as $|\psi|=|\partial\Vec{l}/\partial\ln{r}|$, where $\Vec{l}$ is the unit angular momentum vector \citep{Ogilvie:1999lr}. Applying this transformation to the derivative operators and the velocities, the hydrodynamic equations take the form:
\begin{align}
    &\mathrm{D}v_x - 2 \Omega_0 v_y = -\rho^{-1}(\partial'_x + q\tau \partial'_y + |\psi|\cos({\tau}) \partial'_z)p,\\
    &\mathrm{D}v_y + (2-q)\Omega_0 v_x =  -\rho^{-1}\partial'_y p,\\
    &\mathrm{D}v_z + |\psi|\Omega_0 \sin({\tau}) v_x = -\Omega_0^2 z' - \rho^{-1}\partial'_z p,\\
    &\mathrm{D}\rho = -\rho[(\partial'_x + q\tau \partial'_y + |\psi| \cos({\tau}) \partial'_z)v_x + \partial'_y v_y + \partial'_z v_z],
\end{align}
where $\Vec{v}$ is the velocity vector relative to the prescribed background warping motion \citep{Ogilvie:2013by} and primes denote the warped coordinates. The transformed substantial derivative is:
\begin{equation}
    \mathrm{D} = \partial'_t + v_x \partial'_x + (v_y+q\tau v_x)\partial'_y + (v_z + v_x |\psi|\cos{\tau})\partial'_z.
\end{equation}
\citet{Ogilvie:2013by} consider laminar solutions to the above equations, where all derivatives with respect to $x'$ and $y'$ are omitted and the solution is assumed to be $2\pi$ periodic in $\tau$. This situation is not unrealistic as it models cases where the warp structure changes on length scales larger than the vertical thickness of the disc and on timescales longer than orbital\footnote{While this separation of timescales is strictly not valid for wavelike warps in exactly Keplerian discs, sufficient de-tuning from Keplerian frequency in the local frame can be achieved by a small change in the orbital shear rate $q$, for instance artificially or as a result of a small induced precession frequency \citep{Ogilvie:2013by}.} \citep[but see][for the stability of these solutions and the transition to turbulence]{Ogilvie:2013lw,Paardekooper:2019ns}. They additionally assumed that the velocities induced by the oscillating pressure gradient due to the warp are linear in $z'$ and vanish at $z'=0$, expressed as
\begin{align}
    v_x(z',t') &= u(\tau)\Omega_0 z',\label{equ:scaling1}\\
    v_y(z',t') &= v(\tau)\Omega_0 z',\\
    v_z(z',t') &= w(\tau)\Omega_0 z',
\label{equ:scaling}
\end{align}
where $u$, $v$, and $w$ are dimensionless $2\pi$-periodic functions. The fluid equations in terms of these new velocities then simplify to: 
\begin{align}
&d_{\tau}u + (w + u|\psi|\cos{\tau})u = 2 v +  \frac{c_s^2}{{\Omega_0}^2 h_p^2}|\psi|\cos{\tau}, \label{equ:ode1}\\
&d_{\tau}v + (w + u|\psi|\cos{\tau})v = (q-2)u,\\
&d_{\tau}w + (w + u|\psi|\cos{\tau})w + u|\psi|\sin{\tau} = \frac{c_s^2}{{\Omega_0}^2 h_p^2}-1,\\
&d_{\tau} \ln{H} = u|\psi|\cos{\tau}+w,
\label{equ:ode}
\end{align}
where $h_p$ is the scale height, $H$ is the dimensionless scale height $h_p=Hc_s / \Omega_0$, and $d_{\tau}$ denotes the ordinary derivative d/d$\tau$, since all the spatial derivatives have disappeared after the simplifying assumptions. \citet{Ogilvie:2013by} showed that these equations describe two oscillatory modes. The first one is a linear horizontal epicyclic oscillation due to the warp induced oscillations in the radial pressure gradient. The horizontal velocities in this `sloshing' mode \citep[as coined by][]{Dullemond:2022ng} are proportional to $|\psi|$. The second one is a non-linear vertical oscillation mode due to the variation in the velocity divergence caused by the horizontal oscillations. The vertical velocities caused by this `breathing' mode are proportional to $|\psi|^2$. While the two modes are coupled, in the limit of very small warps \citet{Dullemond:2022ng} derived a solution for the induced horizontal oscillations (neglecting the effects of vertical oscillations), which they then used to derive a general expression for the internal torque, unifying the governing equations for the bending wave and diffusive warp regimes. We note, however, that \citet{Fairbairn:2021ns} showed that the vertical oscillations as well as the coupling play an important role even in the case of small warps, limiting the validity of the unified governing equations where these effects are not taken into account \citep{Dullemond:2022ng}.

\subsection{Adding dust}

In this work, we extend this local approach to include the treatment of dust, which is treated here as a pressureless fluid. Following \citet{Ogilvie:2013by}, we consider laminar flows setting the $x'$ and $y'$ derivatives to zero. By contrast, in this work we do not assume linearity in $z'$ for either the gas or dust component. This assumption is valid for the case of a gas disc since the vertical pressure gradient vanishes at the midplane. However the drag force for the dust and its backreaction will not necessarily vanish at the midplane, making this assumption inappropriate for our coupled gas-dust system. The resulting equations for the gas evolution are

\begin{align}
   \partial'_{t}v_x &+ (v_z + v_x|\psi|\cos{\tau})\partial'_z v_x = 2\Omega_0 v_y \nonumber \\ &- \frac{1}{\rho}|\psi|\cos({\tau})\partial'_z p - \frac{\epsilon(v_x-V_x)}{t_s}, \label{equ:gas1}\\
   \partial'_{t}v_y &+ (v_z + v_x|\psi|\cos{\tau})\partial'_zv_y = (q-2)\Omega_0 v_x - \frac{\epsilon(v_y-V_y)}{t_s},\\
   \partial'_{t}v_z &+ (v_z + v_x|\psi|\cos{\tau})\partial'_z v_z + |\psi|\Omega_0\sin({\tau})v_x = -{\Omega_0}^2 z'\nonumber \\ &- \frac{1}{\rho}\partial'_z p - \frac{\epsilon(v_z-V_z)}{t_s},\\
   \partial'_{t}\rho &+ (v_z + v_x|\psi|\cos{\tau})\partial'_z \rho=-\rho[|\psi|\cos({\tau})\partial'_z v_x+\partial'_z v_z].
\label{equ:gas}
\end{align}
And equivalently for dust:
\begin{align}
   &\partial'_{t}V_x + (V_z + V_x|\psi|\cos{\tau})\partial'_z V_x = 2\Omega_0 V_y - \frac{(V_x-v_x)}{t_s}, \label{equ:dust1}\\
   &\partial'_{t}V_y + (V_z + V_x|\psi|\cos{\tau})\partial'_zV_y = (q-2)\Omega_0 V_x - \frac{(V_y-v_y)}{t_s},\\
   &\partial'_{t}V_z + (V_z + V_x|\psi|\cos{\tau})\partial'_z V_z + |\psi|\Omega_0\sin({\tau})V_x = -{\Omega_0}^2 z'\nonumber \\ &- \frac{(V_z-v_z)}{t_s},\\
   &\partial'_{t}\rho_{d} + (V_z + V_x|\psi|\cos{\tau})\partial'_z \rho_{d}=-\rho_{d}(|\psi|\cos({\tau})\partial'_z V_x+\partial'_z V_z),
\label{equ:dust}
\end{align}
where $V_x$, $V_y$, $V_z$ are the components of the dust velocity, $t_s$ the drag stopping time, $\epsilon$ the dust-to-gas ratio, and $\rho_{d}$ is the dust density. The horizontal and vertical oscillations in the gas component are induced by the radial pressure gradient variations due to the warp. While the dust component does not feel a pressure gradient, it will be affected by these oscillations through the drag force. In the next section we investigate this interaction using a simple numerical 1D (vertical) calculation of Equations \ref{equ:gas1}~-~\ref{equ:gas} and \ref{equ:dust1}~-~\ref{equ:dust}.

\begin{figure*}
    \centering
    \includegraphics[width=\columnwidth]{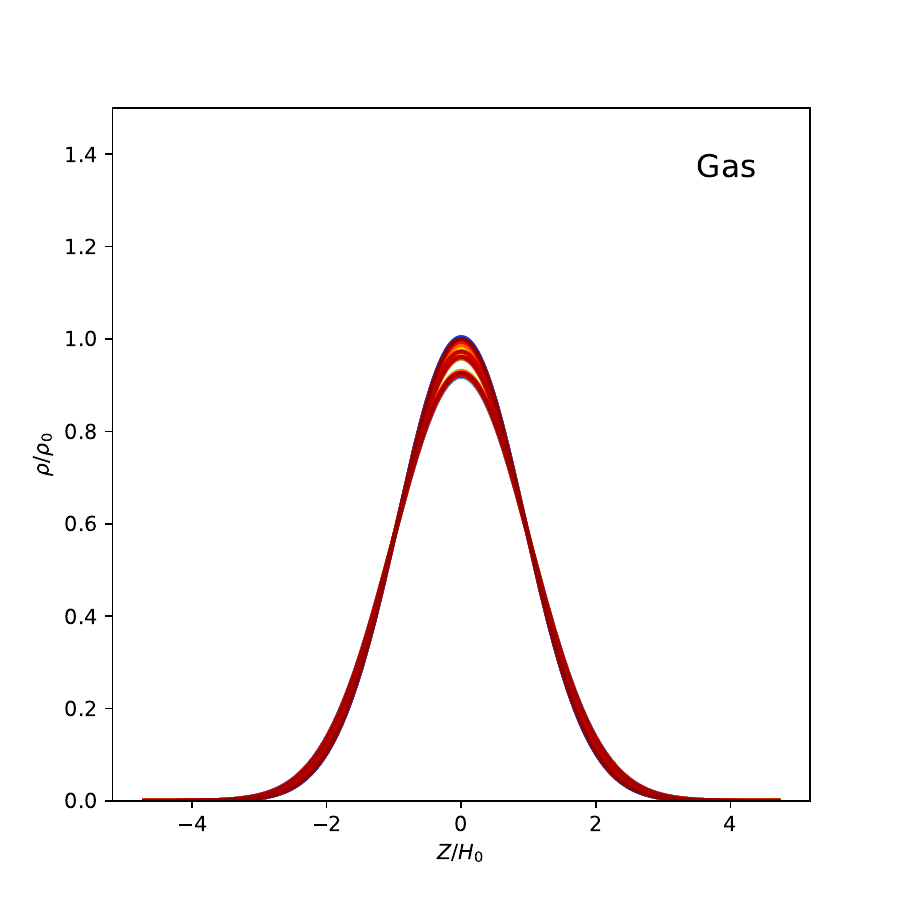}
    \includegraphics[width=\columnwidth]{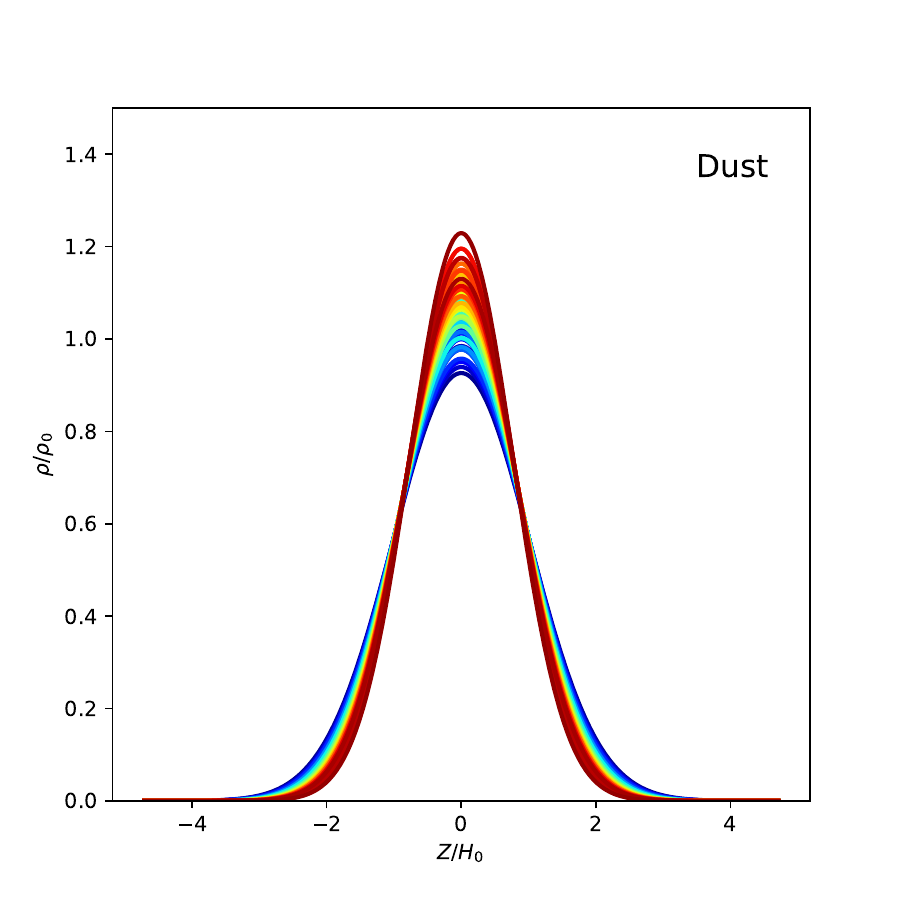}
    \caption{Vertical density profile of the gas (left) and dust (right) for 10 orbits (blue to red). Note that the gas density oscillations are bounded, whereas the dust density increases with time.}
    \label{fig:dens_gas}
\end{figure*}

\section{1D ``Toy Model" Results}
\label{section:1D_results}
We employ the Method of Lines technique and central differencing to discretize the vertical spatial derivatives and effectively transform the PDEs in Equations~\ref{equ:gas1}~-~\ref{equ:gas} and \ref{equ:dust1}~-~\ref{equ:dust} to a series of ODEs, which are to be solved at each grid point. We then use the Python \textsc{solve_ivp} method to solve the initial value problem and advance the solution through time, using an explicit $8^{th}$ order Runge-Kutta scheme. Our solution domain extends from $-5H_0$ to $5H_0$ (where $H_0\equiv c_s/\Omega_0$) and we use open boundary conditions where the fluid properties at the two end grid points in each direction are linearly extrapolated from the preceding two grid points. We discretize our domain to 300 equally-spaced grid points. We set $\Omega_0=c_s=1$ and $\epsilon=0.01$.

The initial conditions for the gas are obtained through evolving the set of ODEs~\ref{equ:ode1}~-~\ref{equ:ode} for 1000 orbits (long enough to ensure the gas oscillations have reached a steady periodic state) and scaling the resulting velocities using Equations~\ref{equ:scaling1}~-~\ref{equ:scaling}, along with the corresponding $h_p$. The initial conditions used for the ODEs ~\ref{equ:ode1}~-~\ref{equ:ode} are chosen to be those of an isothermal hydrostatic equilibrium with no warp. We use a shear rate $q=1.6$ to avoid complications arising from the resonance with the epicyclic frequency \citep[see][]{Ogilvie:2013by}.The dust is initialised with zero velocities and the same $h_p$ as that of the gas. For the dust evolution, we set the $-\Omega_0^2 z'$ to zero as this allows us to represent a fixed thickness dust disc within this 1D Eulerian treatment with initial vertical velocities set to zero. In reality, a dust disc maintains a finite thickness by having a distribution of vertical velocities that balances the vertical gravity component, essentially making each dust particle have a different, slightly tilted orbit (which is challenging to represent in a 1D Eulerian algorithm since each cell represents the average velocity of all particles). This harmonic oscillator behaviour is then damped by gas drag and the dust settles to a thinner disc that is maintained by gas diffusion. Our simplified toy model here neglects this settling and instead focuses on modelling the dust reaction to the warp induced oscillations in the gas. The Stokes number St, defined as $\mathrm{St}=\Omega_0 t_s$, is related to the gas density in the Epstein drag regime \citep{Epstein:1924vw} through 
\begin{equation}
  \mathrm{St}=\frac{\pi}{2}\frac{\rho_s s}{\Sigma_g}  
\end{equation}
where $\rho_s$ is the dust intrinsic density, $s$ is the dust particle size, and $\Sigma_g$ is the gas surface density. Our simplified model here fixes St per calculation, even as the gas density increases towards the mid-plane. The SPH simulations Section~\ref{section:3D_results} naturally takes all of this into account.

\begin{figure*}
	\includegraphics[width=\textwidth]{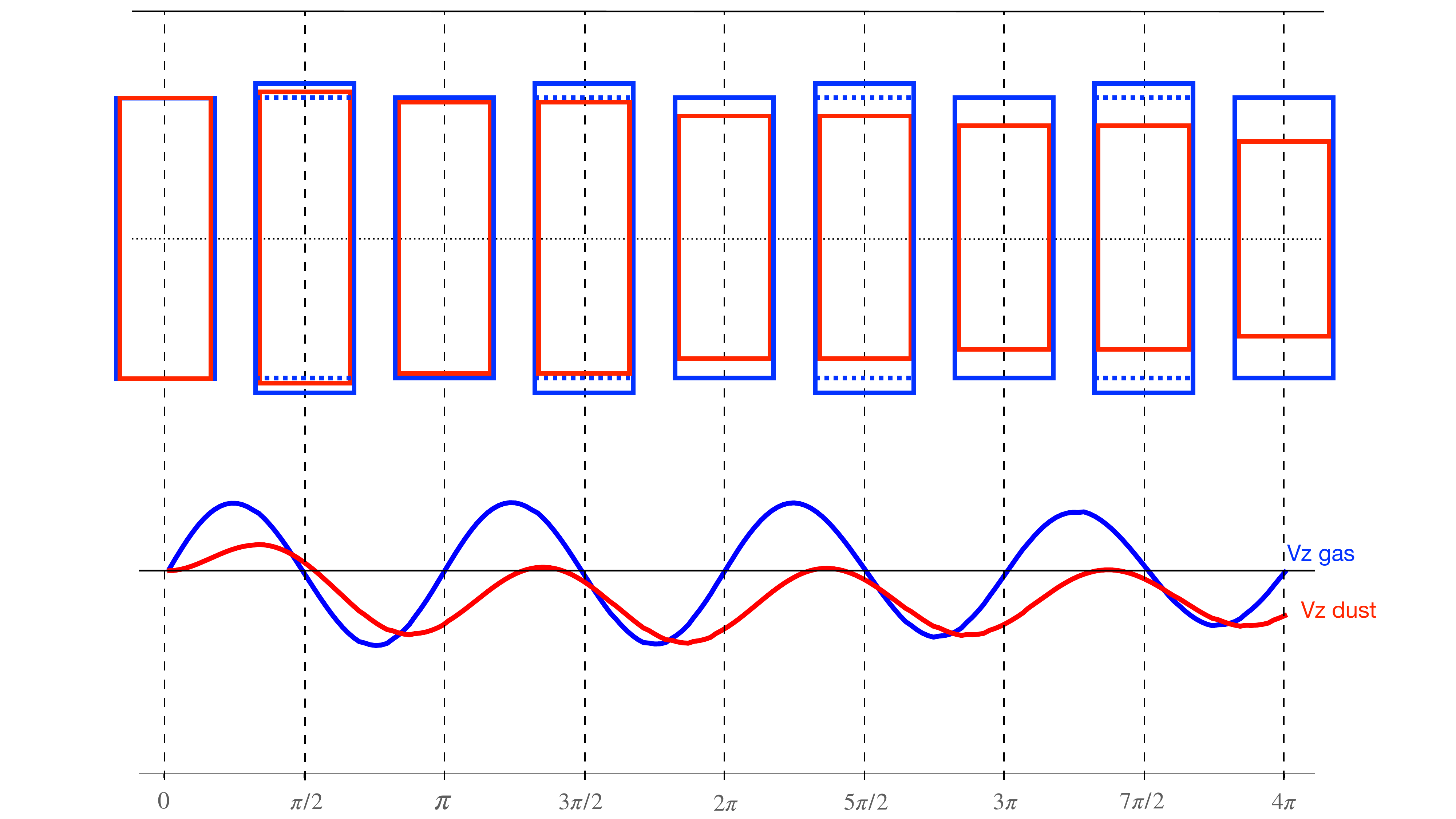}
    \caption{A schematic showing the local geometry of the breathing modes for the gas and dust in a warped disc. The top boxes are scaled based on the $H$ values obtained from a 1D calculation with St$=1$ for both gas (blue) and dust (red), with the blue dotted box indicating the initial vertical thickness (equal in both gas and dust). The boxes are plotted every $\pi/2$ for $2$ full orbits. The bottom panel shows the gas (blue) and dust (red) vertical velocities \textit{measured at the upper boundary of the dust box shown in the upper panel}, demonstrating the velocity difference acting on the dust element, as well as the resulting phase offset. Note that we used St$=1$ so that the effect in the upper panel is strong enough to be easily visible. }
    \label{fig:schematic}
\end{figure*}

The left panel of Figure~\ref{fig:dens_gas} shows the gas vertical density profiles calculated by our method at different times, progressing from blue to red, for a total of 10 orbits (plotting 5 times per orbit at equal intervals). For this fiducial calculation we use dust with St=$0.1$, and a warp magnitude $|\psi|=0.1$. We see that the gas density profile oscillates (see Figure~\ref{fig:schematic} for a clear illustration of the breathing mode responsible for these oscillations), but without any discernible cumulative effects, as the bounds of the oscillations are identical and the warm curves (later times) cover the cool curves (earlier times).

By contrast, we show the corresponding profiles for the dust component in the right panel of Figure~\ref{fig:dens_gas}. Here we see that in addition to the oscillations experienced by the gas component there is a clear cumulative effect resulting in a net compression of the dust. The physical interpretation of this is the core of the instability: The breathing mode induced in the dust component is only driven by the drag force exerted by the gas, which is a function of the velocity difference between dust and gas. The relevant gas velocity is expected to be approximately linear in altitude $z$, which means a dust fluid element, for example above the mid-plane, would feel a greater downward velocity from the oscillating gas further up (in a contraction stroke) than the upward oscillating gas further down (in the subsequent expansion stroke). This causes a net compression effect on the dust, as well as inducing a phase offset between gas and dust, which is demonstrated schematically in Figure~\ref{fig:schematic}. Because this effect relies on the coupling between gas and dust, increasing the decoupling by for example, increasing the St, will enhance this effect. We expect a moderate St to be optimal for dust concentration, where the growth rate is larger but the dust is not yet too decoupled to prevent WInDI from occurring. We also note that this net compression effect would be partially counteracted by a vertical dependence of St (not implemented in this 1D toy model), as well as the turbulent diffusion caused by settling effects. On the other hand, it will be enhanced by the effects of vertical gravity (turned off here).

% Schematic figure

\subsection{Varying dust size}

\begin{figure}
    \centering
    \includegraphics[width=\columnwidth]{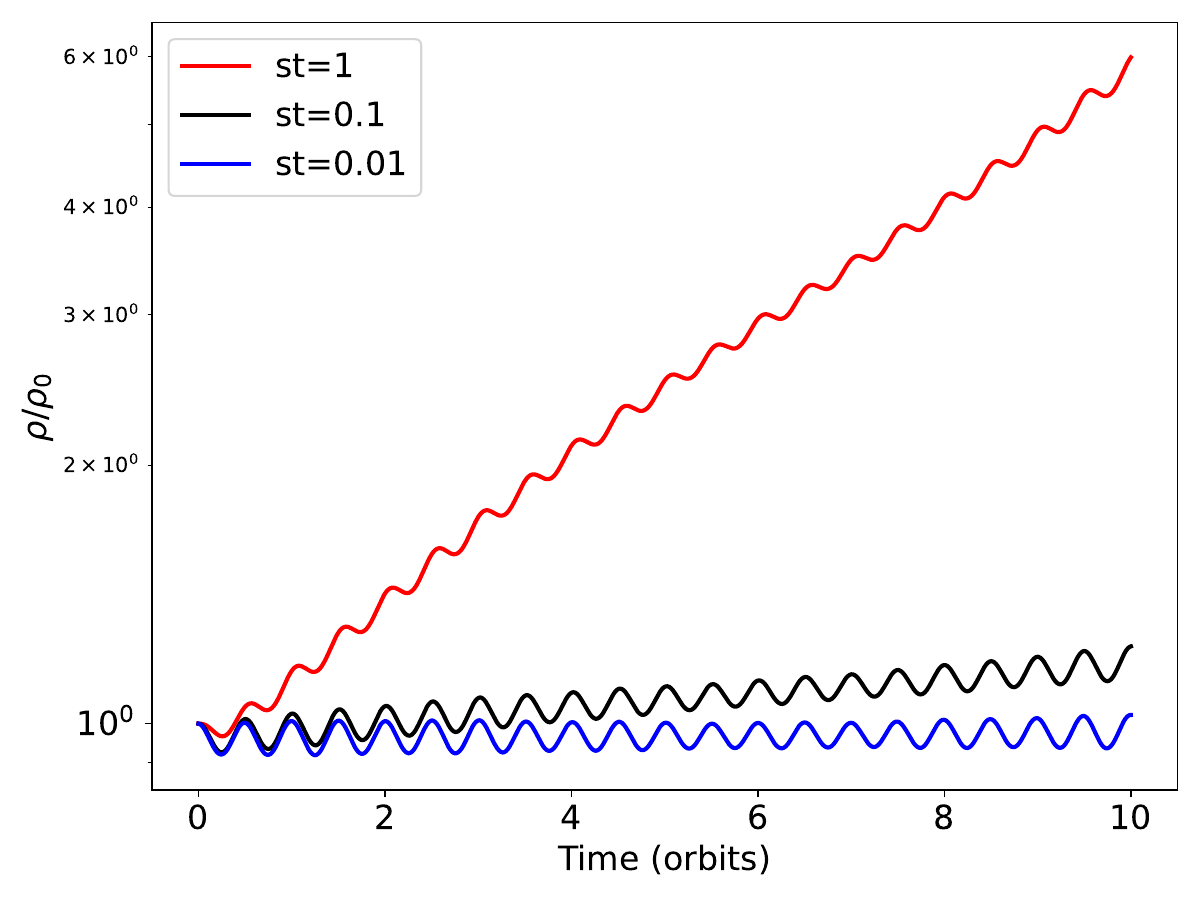}
    \caption{Time evolution of the midplane dust density as a function of St and a fixed warp amplitude of $|\psi|=0.1$. As predicted from the 1D framework, the instability is more prominent for higher St but the amplitude of the oscillations (driven by the gas) are independent of St.}
    \label{fig:trend_st}
\end{figure}

In Figure~\ref{fig:trend_st} we show the midplane dust density evolution as a function of time for three different dust sizes, corresponding to St$=0.01$, $0.1$, and $1$, and a fixed warp amplitude of $|\psi=0.1|$. The dust density grows exponentially at a rate of $\sim 1.25$/orbit for the St=$1$ case and thus shows signs of an instability. As expected from our physical interpretation presented above, dust with higher St (ie, more decoupled from the gas) shows a greater cumulative compression with time. The more decoupled dust particles slide further from the gas in each breathing cycle, resulting in greater asymmetry between the compression and de-compression strokes, and thus a stronger net compression.

We note that, while the cumulative compression increases with dust size (St is proportional to the grain size in the Epstein drag regime), the magnitude of the oscillations is similar in all three cases. This is due to the fact that the oscillations are driven only in the gas and the dust only responds to them, hence the oscillations magnitude is independent of St.

\subsection{Varying the warp amplitude}
\begin{figure}
    \centering
    \includegraphics[width=\columnwidth]{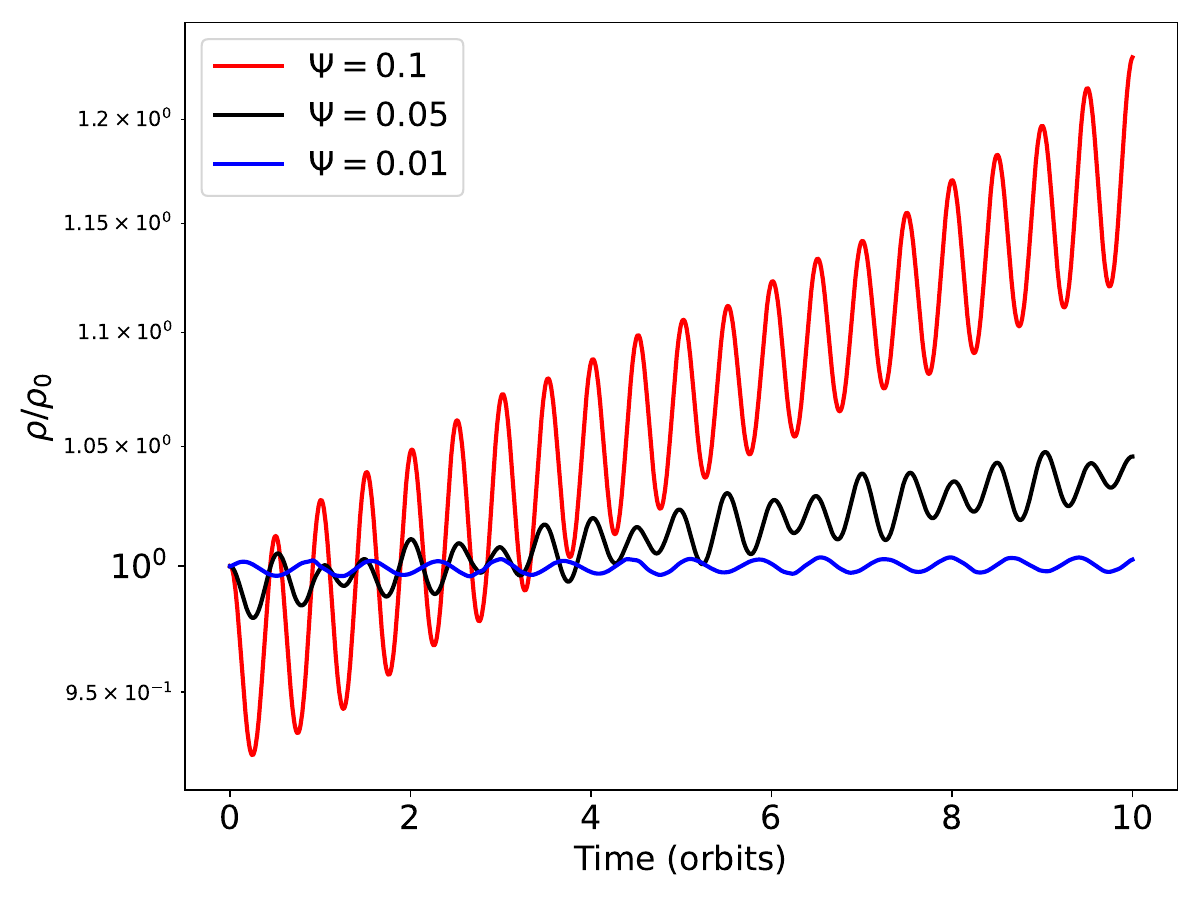}
    \caption{Time evolution of the midplane dust density as a function of warp amplitude and a fixed dust size corresponding to St=$0.1$. As expected from the 1D framework, the instability is enhanced for larger warp amplitudes as well as the amplitude of the oscillations.}
    \label{fig:trend_psi}
\end{figure}

Figure~\ref{fig:trend_psi} shows the time evolution of the dust midplane density for three different warp amplitudes $|\psi|=0.01$, $0.05$, and $0.1$, and a fixed dust size corresponding to St=$0.1$. We see that the dust midplane density increases much quicker with stronger warps, as expected. We also note that both the net compression and the oscillation magnitude increase for stronger warps, as the warp affects the breathing mode in the gas (as opposed to varying St in Figure~\ref{fig:trend_st}).

\section{Global SPH simulations}
\label{section:3D_results}
Our simplified 1D analysis suggests that the warp-induced oscillations in the gas triggers an instability in the dust component that gets stronger with St and $|\psi|$. However, two main questions remain unanswered: first, whether this instability has a global manifestation or is only a local phenomenon, and second, how is it affected by the warp evolution.

We tackle these issues by performing 3D global SPH simulations of a warped gas and dust disc with varying St. SPH has been widely utilised in simulating disc warps as it has intrinsic advantages in this regime over grid-based numerical schemes   \citep[although they have been used, e.g.][]{Fragner:2010rt,Deng:2022lj,Rabago:2023jw}. The most important of these is the fact that SPH has no preferred directions that could influence warp alignment as well as its Galilean-invariance resulting in its numerical dispersion being independent of any flow velocity.

We use the SPH code \textsc{Phantom} \citep{Phantom}, which offers two methods for modelling gas and dust mixtures. The first is a 2-Fluid algorithm which treats the gas and dust components separately, solving a different set of governing equations for each component and computes a drag coupling term \citep{Laibe:2012vu,Price:2020af}. This method computes the stopping time explicitly and the time stepping is chosen so that the drag interaction is time-resolved (making this method relatively computationally expensive). The second method is a 1-Fluid algorithm where one set of equations is solved for the gas-dust mixture, including an evolution equation for the dust fraction \citep{Price:2015bh}. This method employs a terminal velocity approximation and thus does not resolve the drag interaction. Our intuition, which is based on the results obtained in Section~\ref{section:1D_results}, suggests that the 2-Fluid approach is more appropriate here since the cumulative dust compression relies on an appropriate time-resolution of the drag interaction through each breathing cycle. Therefore, all the SPH simulations presented here employ the 2-Fluid algorithm. We note that equivalent 1-Fluid simulations (not shown) did not recover WInDI with the parameters used in this work - as expected.

\subsection{Setup}
We set up a disc with inner and outer radii $R_{\mathrm{in}}=10$~au and $R_{\mathrm{out}}=100$~au around a star of mass $M_*=1M_\odot$. The central star is modelled as a sink particle with an accretion radius of $R_{\mathrm{acc}}=5$ au. The gas disc has mass $M_{\mathrm g}=10^{-3}M_\odot$ represented by $10^6$ equal mass SPH particles distributed such that the initial surface density profile obeys the power law:
\begin{equation}
    \Sigma_{\mathrm g}=\Sigma_{\mathrm g,0}\left(\frac{R}{R_{\mathrm{in}}}\right)^{-p},
\end{equation}
where $\Sigma_{\mathrm{g},0}$ is a normalisation constant and $p=1$. We use a globally isothermal equation of state and the particles are distributed vertically according to a Gaussian profile with an aspect ratio $\frac{H}{R}=0.14$ at the warp radius $R_{\mathrm w}=50$ au. We use a constant SPH artificial viscosity that corresponds to a viscosity coefficient \citep{shakura_sunyaev} of $\alpha=0.01$. In practice, the resulting effective viscosity is usually somewhat larger. While we do not attempt to directly measure the effective dissipation in our simulations, we note that \citet{Aly:2021bu} have indirectly estimated a viscosity coefficient of $\alpha=0.05$ for SPH simulations with similar parameters, by means of comparing the long term evolution with that of a 1D ring code. This puts us safely in the bending wave warp regime. All times are quoted in units of $P_{\mathrm{orb}}$, the orbital period at the warp radius.

We let the gas disc relax for $5$ orbits at $R_{\mathrm{out}}$ before we add the dust component to make sure we do not develop spurious dust features due to the initial random placement of gas particles. The dust component is represented by $2 \times 10^5$ SPH particles with mass $M_{\mathrm d}=10^{-5}M_\odot$ with the same radial extent and surface density scaling as that of the gas. However, we choose the thickness of the dust component to be five times smaller than that of the gas in order to minimise any effects caused by dust settling \citep[eg., the settling instability,][]{Squire:2018bb}. We run simulations with $4$ different dust sizes of (100{\textmu}m, 1mm, 1cm, 1m) which correspond to an average St of (0.1, 1, 10, 100). We set up an initial warp at radius $R_{\mathrm w}$ with a width of $30$ au and inclination of $30^{\circ}$ following the procedure outlined in \citet{lodato_2010}. The tilt is described with an increasing sinusoidal profile between $20$ au and $80$ au, with a corresponding maximum warp amplitude of $|\psi|=0.2$.

\subsection{Global WInDI}
\begin{figure}
  \begin{center}

    \includegraphics[width=42.5mm]{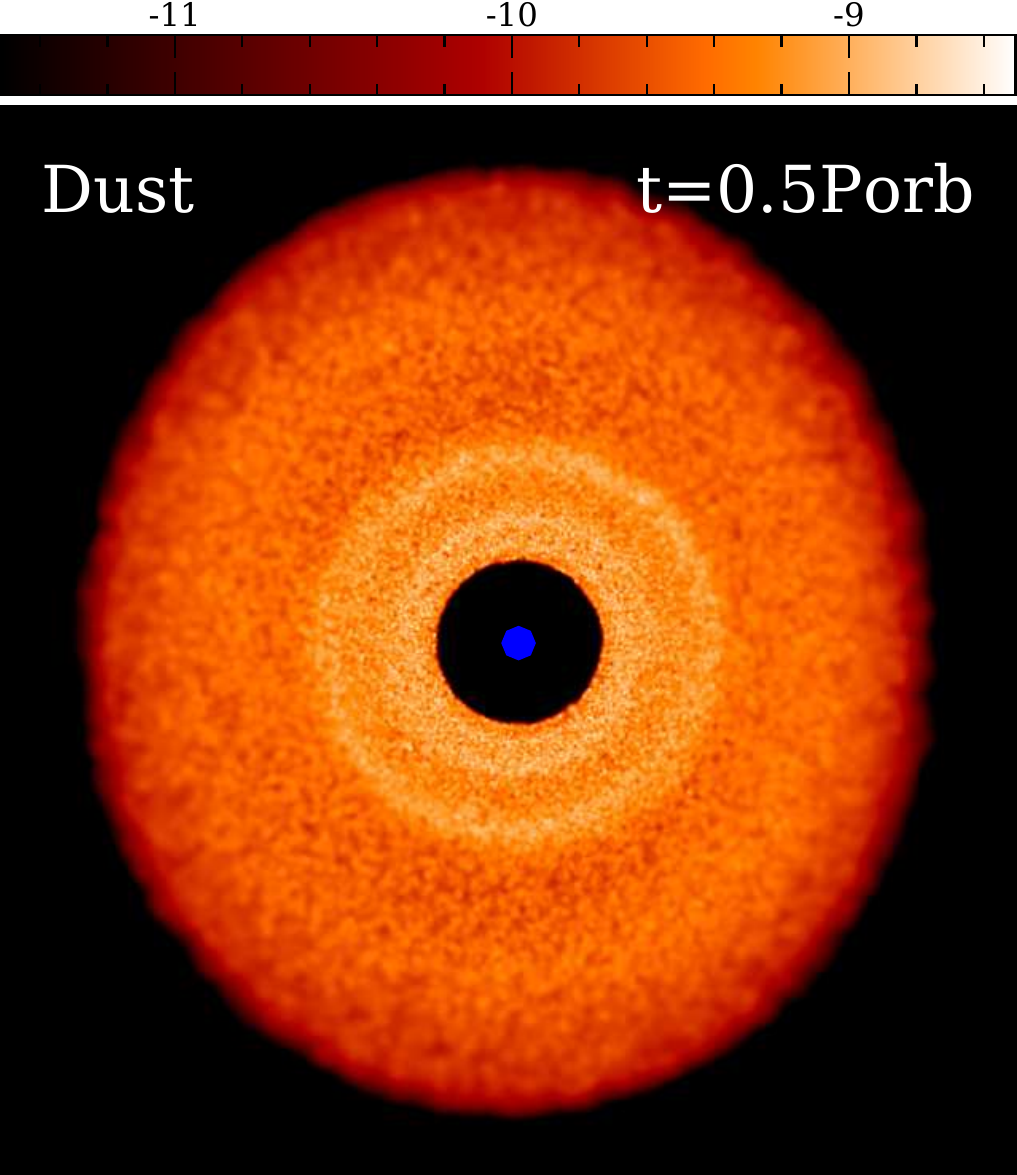} \hfill
    \includegraphics[width=42.5mm]{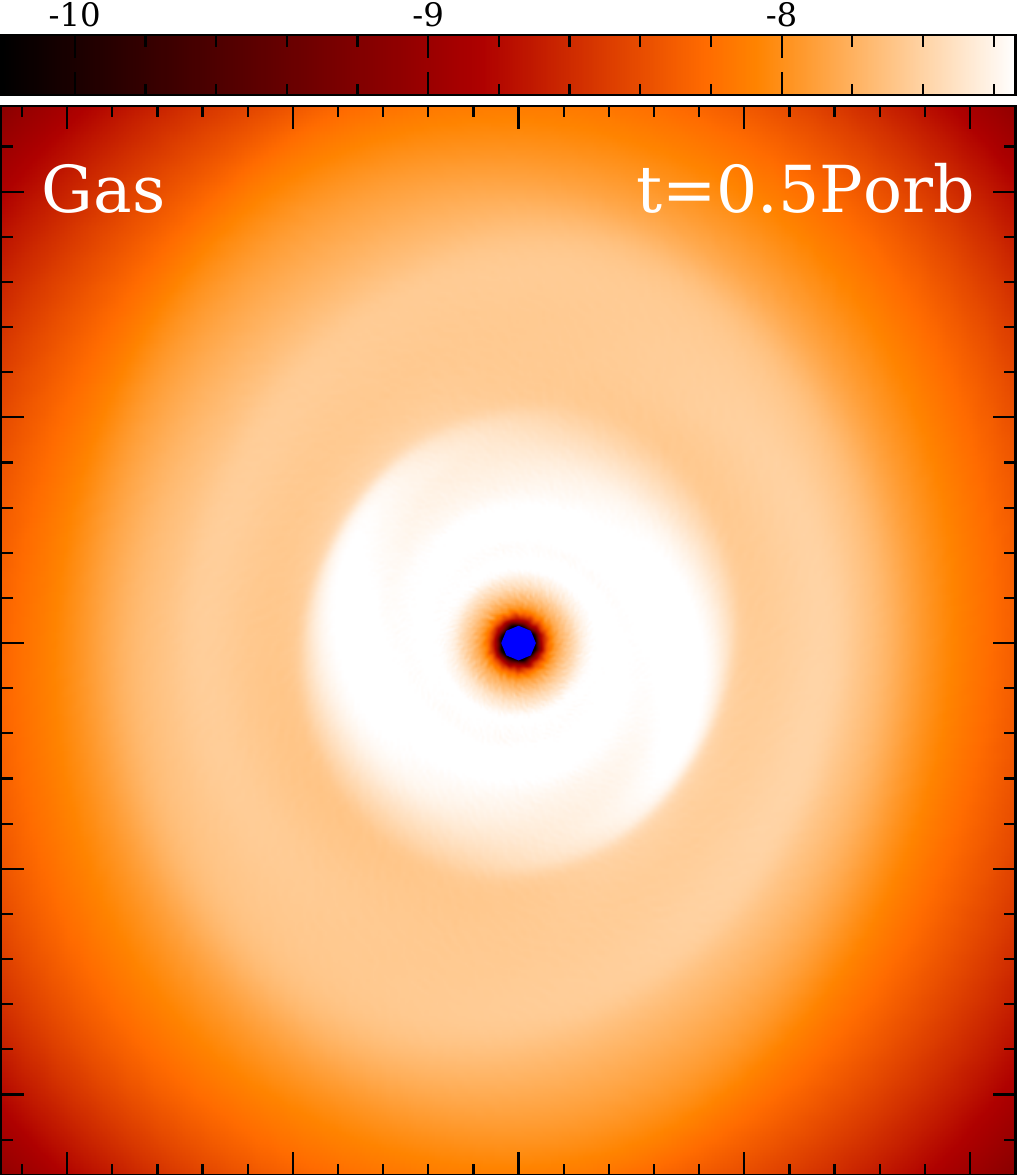}

    \includegraphics[width=42.5mm]{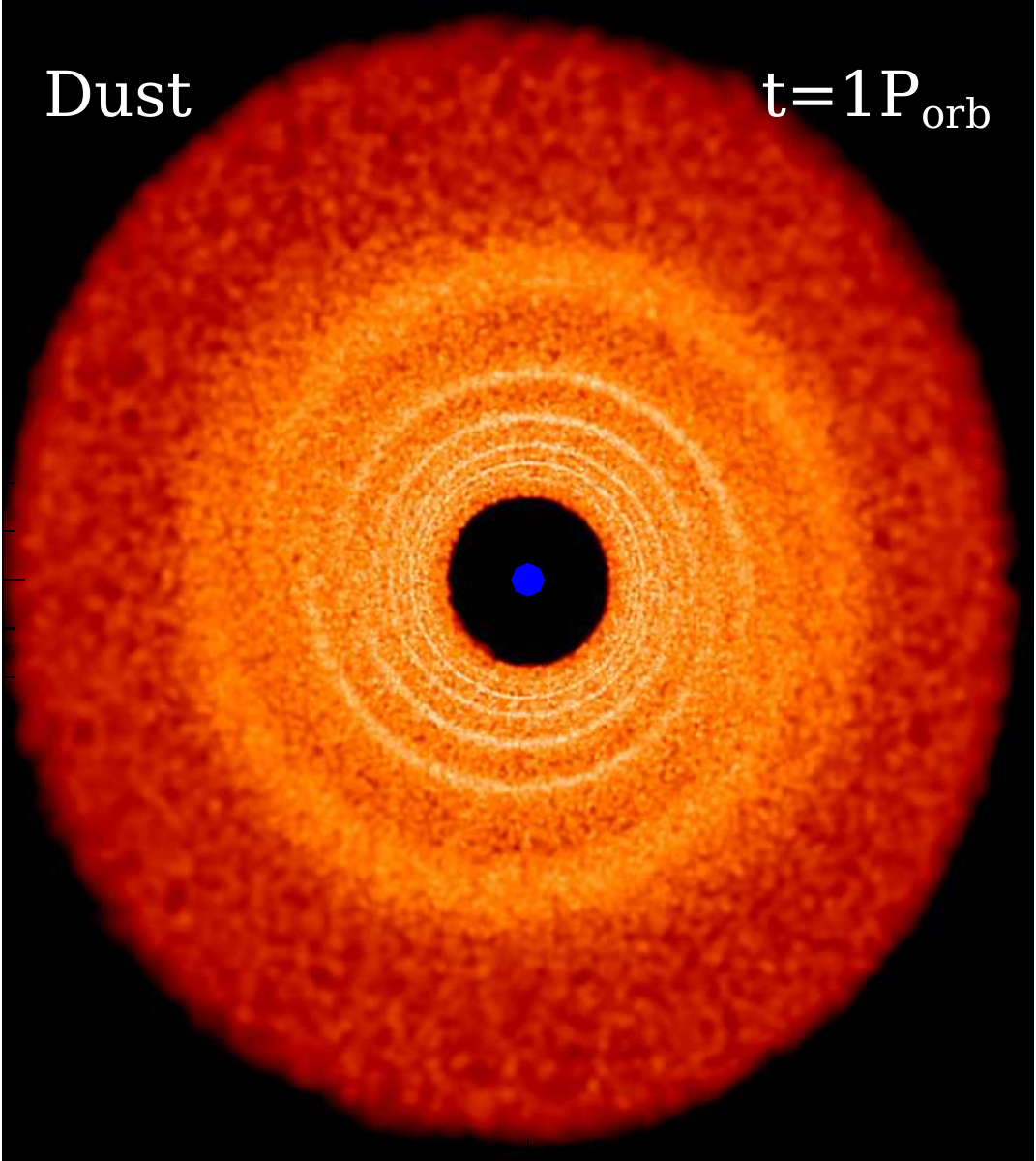} \hfill
    \includegraphics[width=42.5mm]{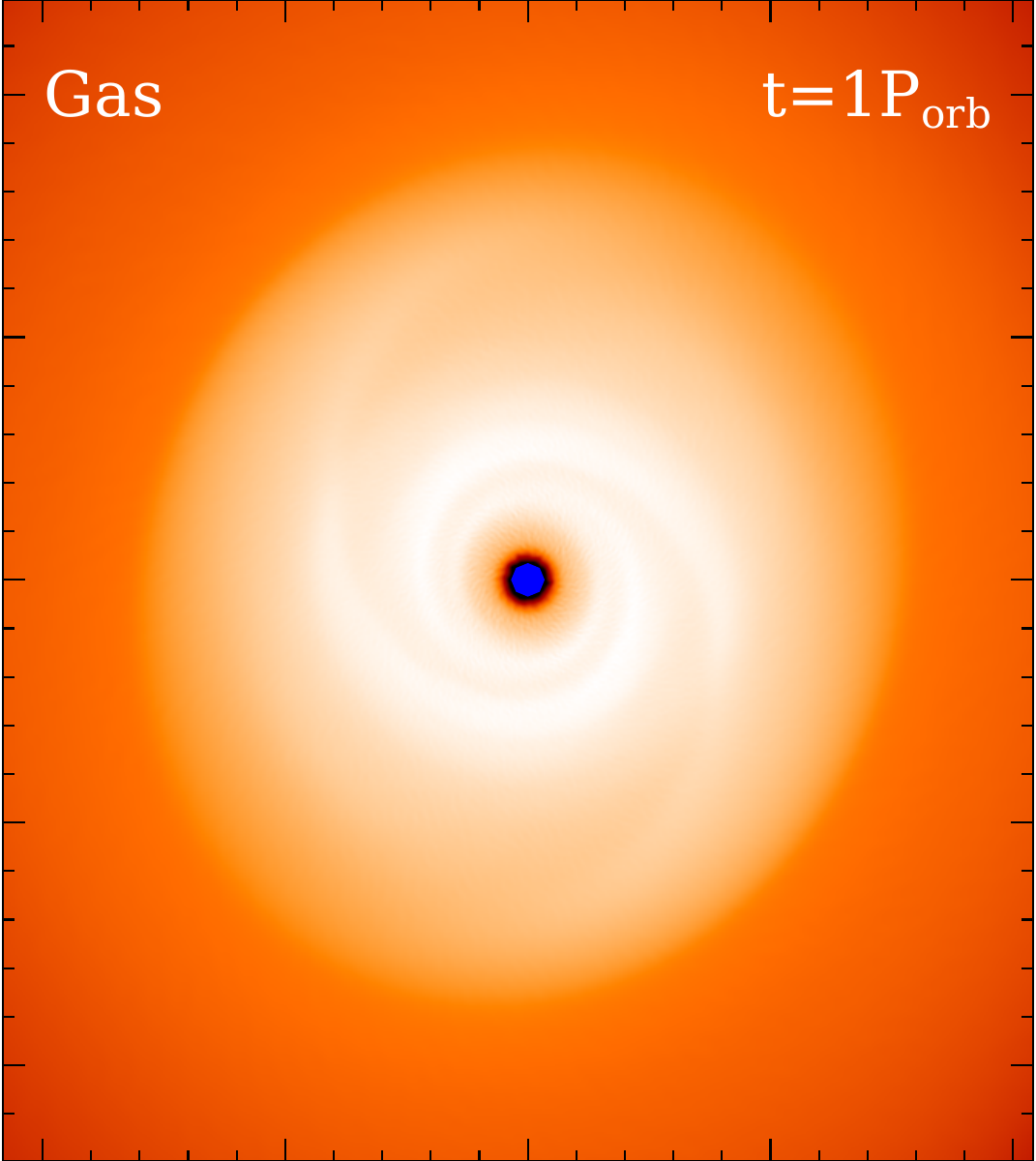}

    \includegraphics[width=42.5mm]{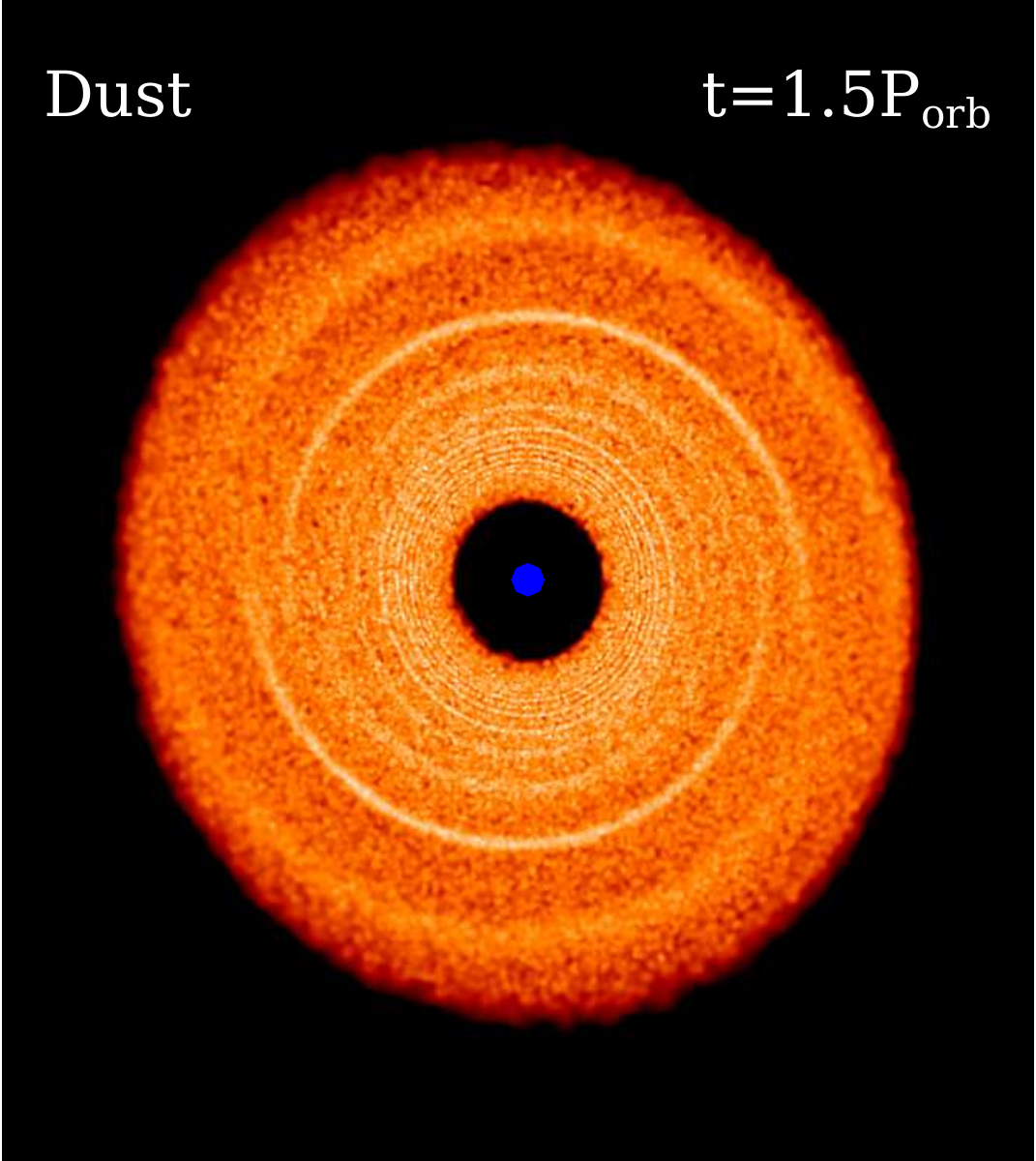} \hfill
    \includegraphics[width=42.5mm]{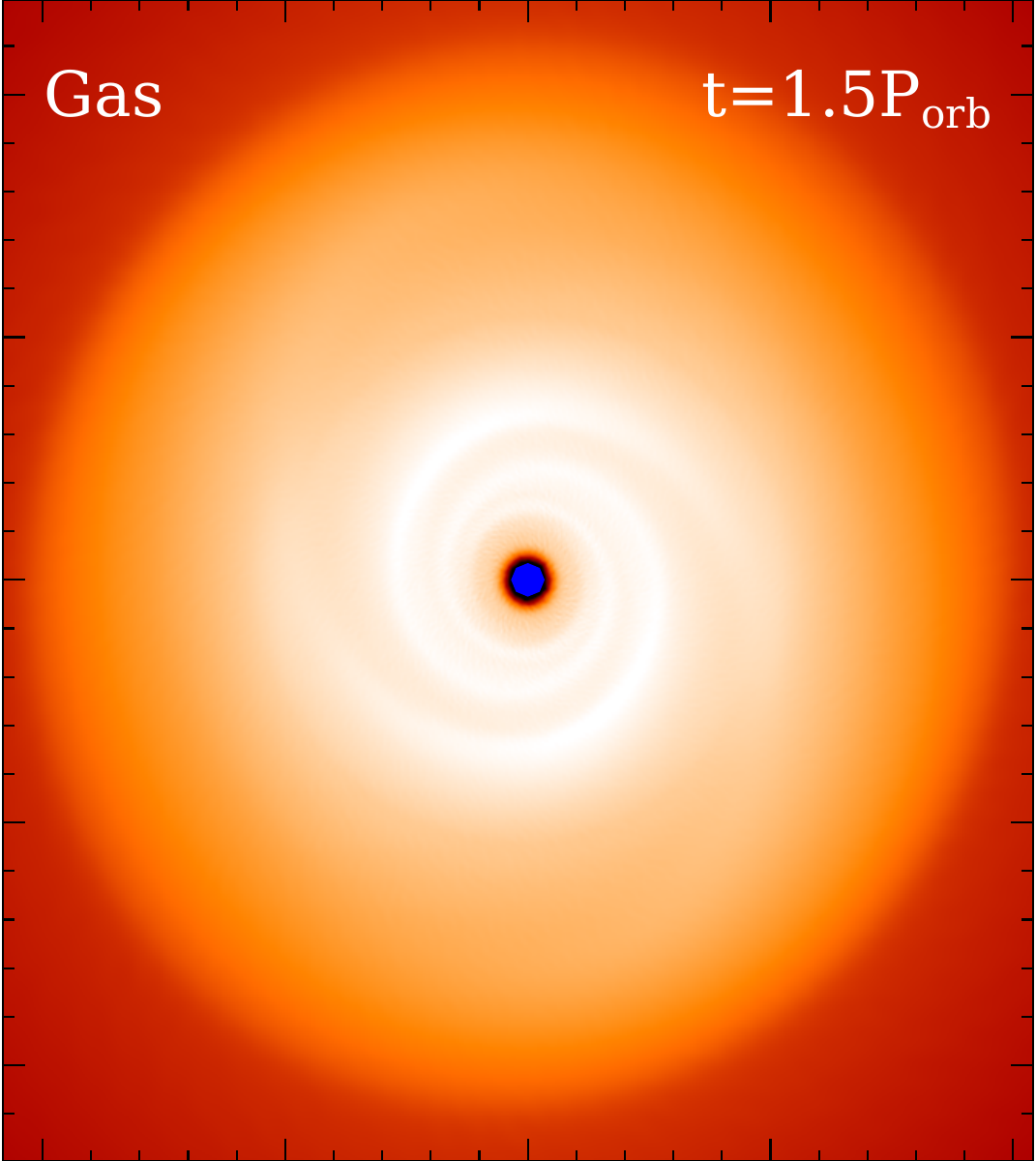}

    \includegraphics[width=42.5mm]{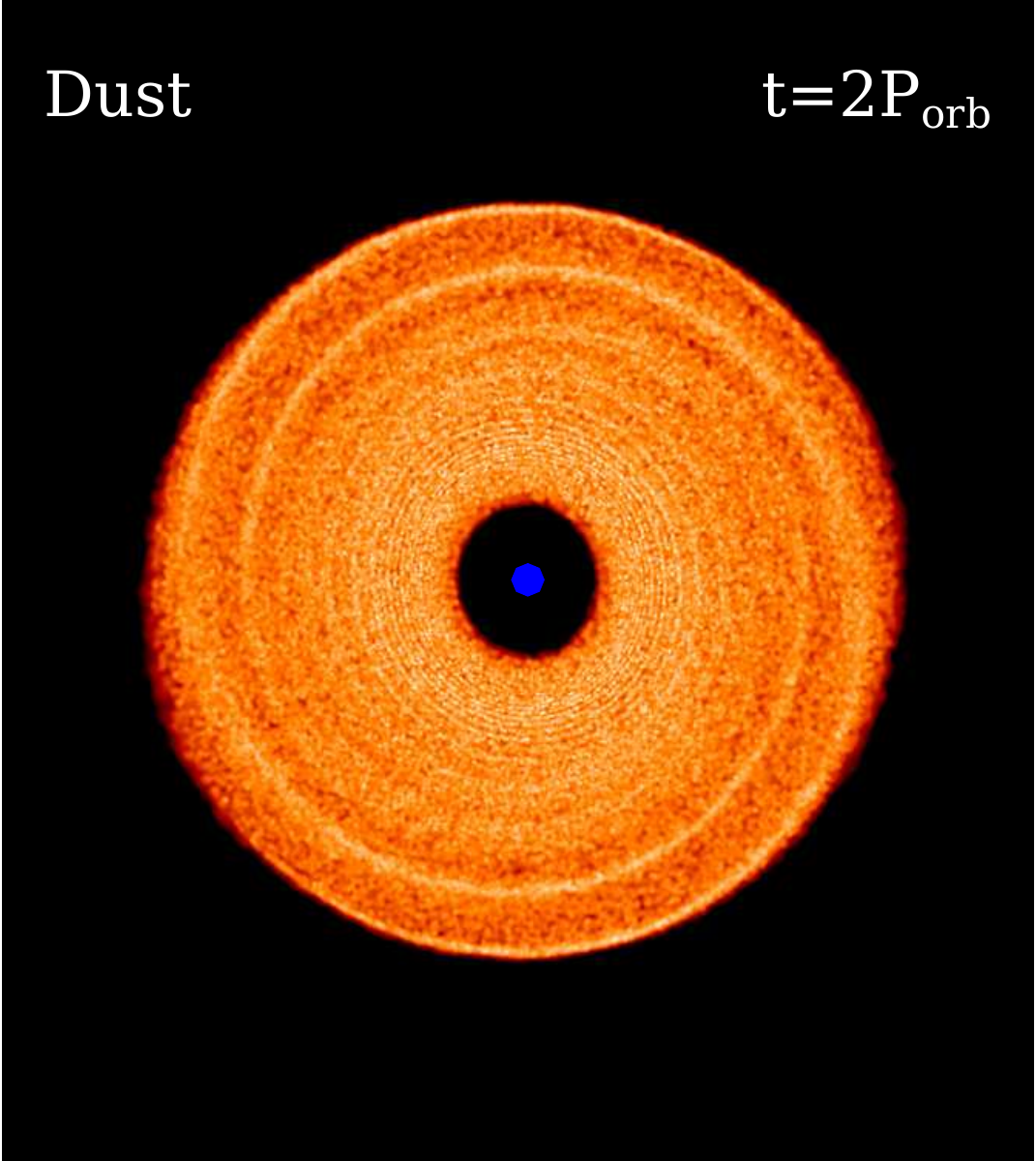} \hfill
    \includegraphics[width=42.5mm]{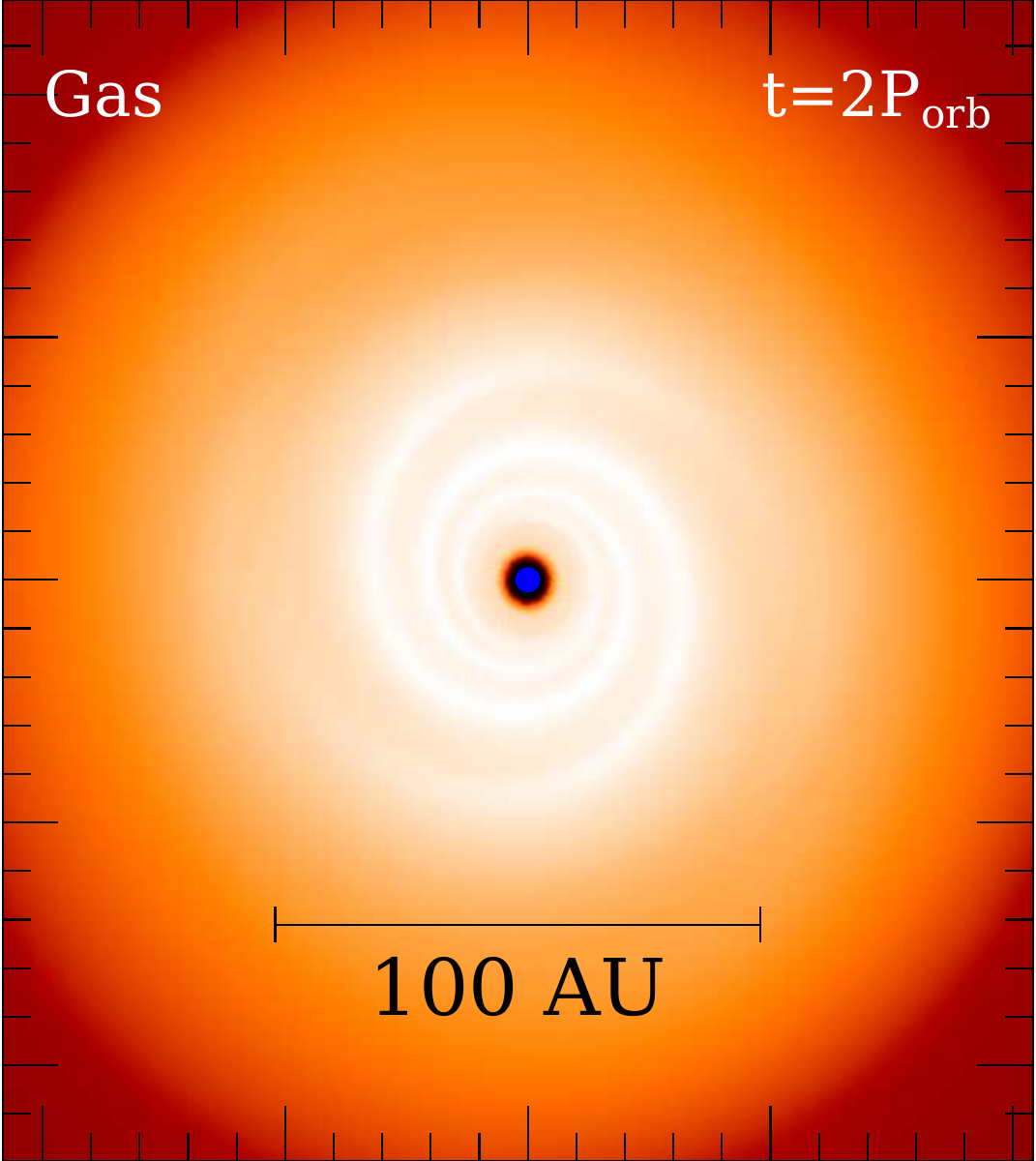}
 \caption{Time evolution of the column density (logarithmic in M$\odot$/AU$^2$) in the X-Y plane for both dust and gas at different times for the St $\sim 0.1$ case.
 \label{fig:splash_time_evolution}}
 \end{center}
\end{figure}

The time evolution for the column density of both dust and gas for the St $\sim 0.1$ case is shown in Figure~\ref{fig:splash_time_evolution}. We can see that WInDI manifests globally as the formation of dust structures. The local dust compression due to the mechanism explained in Section~\ref{section:1D_results} spreads throughout the disc due to the radial motions induced by the warp, as well as the propagation of the warp through the disc. Our 3D calculation suggests that WInDI has two key ingredients, the vertical compression identified by our 1D toy model and the sloshing effect driven by the presence of a warp which together cause radial concentrations in the dust. The importance of the sloshing is evidenced in the broken nature of the rings: where the sloshing effect due to the warp is minimal the dust concentration is neglible but one quarter of an orbit later where the sloshing effect is maximised, the dust ring has the highest concentration. Additionally, the slight offset across the break corresponds to the change in direction of the slosh. Analogous to the role of the breathing mode illustrated by our toy model in section~\ref{section:1D_results}, the sloshing motion has a vertical dependence which produces a net horizontal push as the dust height varies during an oscillation cycle, leading to dust density enhancement at certain locations.

Importantly, we see from Figure~\ref{fig:splash_time_evolution} that the dust structure does not exactly correspond to the gas spirals\footnote{These gas spirals are transient features caused by the introduction of an idealised warp to an otherwise relaxed disc.} induced by the pressure variations due to the warp. In conjunction with our 1D framework, this confirms that WInDI is not simply due to the trapping of dust at gas pressure bumps produced by the warp profile.

\subsubsection{Measuring the growth rate}
\label{section:measuring_dust_growth}
To measure the growth rate of dust concentrations in our 3D simulation ideally we would isolate the warped dust mid-plane, identify dust particles within a certain distance of this mid-plane, azimuthally average their density to find the warped radial dust density profile and find the maxima of this profile. Finally we would repeat this for all of our simulation snapshots and would then be able to identify the regions where growth is occurring. While this method is likely to capture WInDI in the most suitable way to compare to our 1D model, it also captures the increasing dust density as dust settles to the mid-plane. This introduces ambiguity into our measurements as it is difficult to distinguish concentrations due to dust settling rather than due to WInDI.

Instead, we consider the dust surface density profile as this is more sensitive to the formation of radial dust rings rather than settling perpendicular to the mid-plane. Another contributor to dust concentrations is the inward radial drift caused the headwind felt by dust particles \citep{Whipple:1972bj,Weidenschilling:1977bw}. To address this, in Appendix~\ref{section:dustgrowthflatdisc} we apply this method to measure radial dust concentrations in a flat, non-warped disc, where WInDI does not operate and only the effects of radial drift are applicable.

Following \citet{lodato_2010} we discretise the dust particles of the disc into concentric spherical shells, allowing for the warped mid-plane. We then further discretise the disc into two semi-circles based on the particle's $y$ position. This allows us to construct dust surface density profiles as a function of radius for the two halves of the disc separated at the line of nodes of the warp. Separating the disc into two semi-circles is necessary because of the unique geometry of the rings characteristic of WInDI, where the rings vanish at a certain phase and have different radii between the upper and lower part of the disc. For each radial dust surface density profile we identify the local maxima with the uncertainty the width of the bins in our discretisation process.

\begin{figure*}
    \centering
    \includegraphics[width=\textwidth]{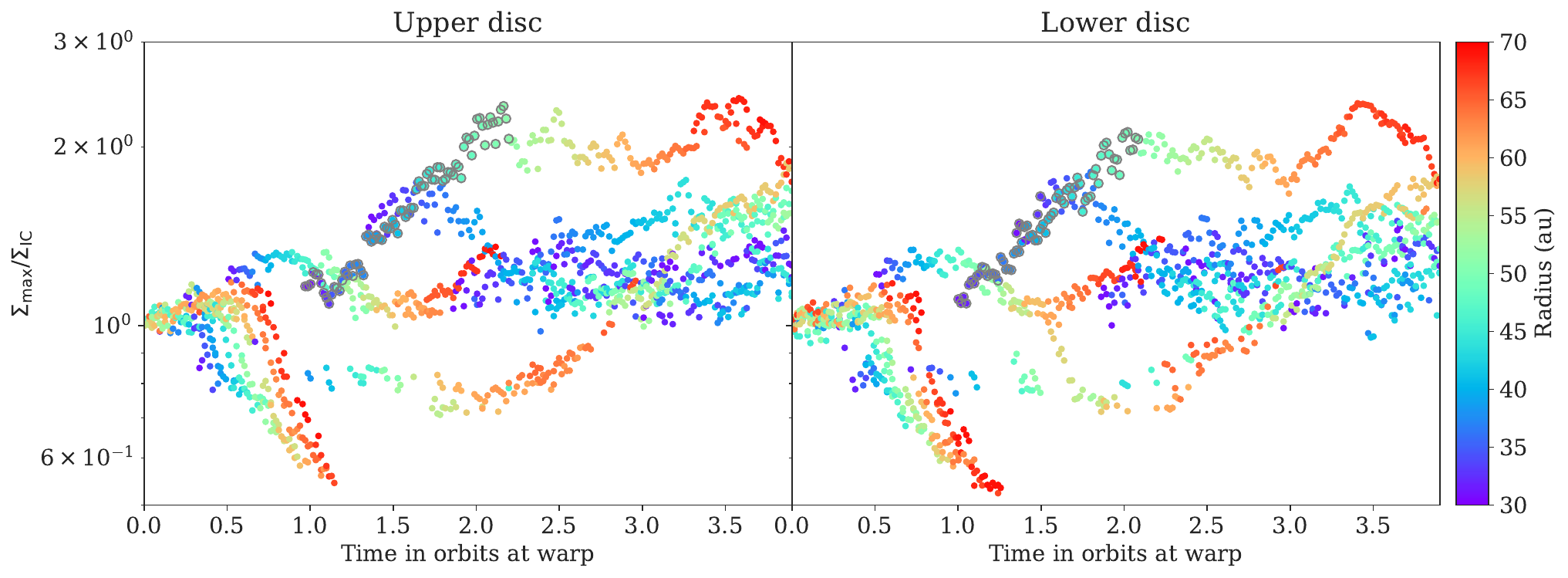}
    \caption{Growth of local dust concentrations throughout the simulation shown in Fig~\ref{fig:splash_time_evolution}. Each point represents a local maximum in the radial dust surface density profile scaled by the initial surface density at that radius. Each point is coloured by its radius, with the uncertainty in the radial measure $\pm1$au. This is shown for the two halves of the disc separately to respect the unique offset ring structure of WInDI. The points circled in grey are associated with the most prominent ring in Fig~\ref{fig:splash_time_evolution} and used to estimate a growth rate of $1.09\pm0.06$/orbit.}
    \label{fig:growth_rates}
\end{figure*}

Figure~\ref{fig:growth_rates} shows the peaks the in dust surface density profiles over the evolution of the simulation shown in Figure~\ref{fig:splash_time_evolution}. To demonstrate relative growth or decay the surface density values are scaled by the initial surface density at that given radius and are coloured by their radial value. For both halves of the disc decay is prominent before $t=1$ orbit at $R_{\mathrm w}=50$ au.
Looking to  Figure~\ref{fig:splash_time_evolution} this is likely due to transients introduced with the dusty disc and warp and does not correspond to features attributed to WInDI. Between $t=1$ and $t=2$ orbits growth is apparent for a ring that starts at $30$au and travels to $50$au (these points are selected with grey circles). We associate this with the second-most outer prominent ring seen in Figure~\ref{fig:splash_time_evolution}. The dust density in this prominent ring broadly plateaus after $t\sim2$ orbits as radial drift at the outer edges begins to dominate the dust evolution in the outer disc (evidenced by the outer edge of the dust disc decreasing in the last panel of Figure~\ref{fig:splash_time_evolution}). 
In agreement with Figure~\ref{fig:splash_time_evolution}, Figure~\ref{fig:growth_rates} also shows dust density growth of a second ring starting at $30$au, $t=1.4$ orbits as well as rings at $60$au, $t=1.5$ orbits (this is the outermost ring in Figure~\ref{fig:splash_time_evolution}) and $60$au, $t=2.0$ orbits.

We measure the growth rate of WInDI in our 3D simulation from the prominent ring that starts at $30$au and $t=1.0$ orbits as it the clearest example. The local warp strength that coincides with this ring has $|\psi|\sim0.2$ but from Section~\ref{section:1D_results}, the largest warp strength examined was $|\psi|=0.1$ in Figure~\ref{fig:trend_psi}. We would then anticipate a growth rate of more than $\sim 1.2$/orbit in our 3D simulations as they have a larger warp. Using a least squares fit with the points on Figure~\ref{fig:growth_rates} indicated with grey circles we measure the growth rate for the upper disc as $1.144\pm0.004$/orbit and for the lower disc as $1.027\pm0.007$/orbit. This gives a combined growth rate of $1.09\pm0.06$/orbit, notably lower than the prediction from the 1D model. We address this discrepancy in Section~\ref{section:limitations}. Comparison between Figure~\ref{fig:growth_rates} and the corresponding growth rates obtained for an, otherwise identical, unwarped disc (and thus growth is only due to radial drift) in Appendix~\ref{fig:flat_growth_rates} shows that the growth rates due to WInDI are distinct from (and higher than) those resulting global radial drift.

\subsubsection{With different St}
\begin{figure*}
  \begin{center}
    \resizebox{175.0mm}{!}{\mbox{\includegraphics[angle=0]{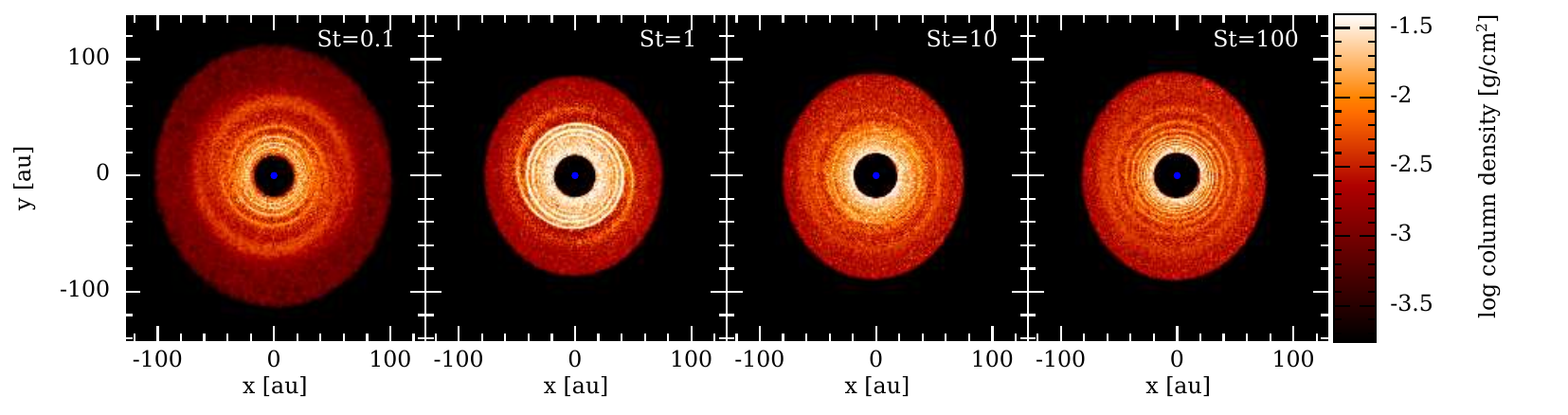}}}
 \caption{Column density of the dust component after 1 orbit at the warp radius for 4 different St.\label{fig:splash_st}
 }
 \end{center}
\end{figure*}

\begin{figure}
  \begin{center}
    \resizebox{78.0mm}{!}{\mbox{\includegraphics[angle=0]{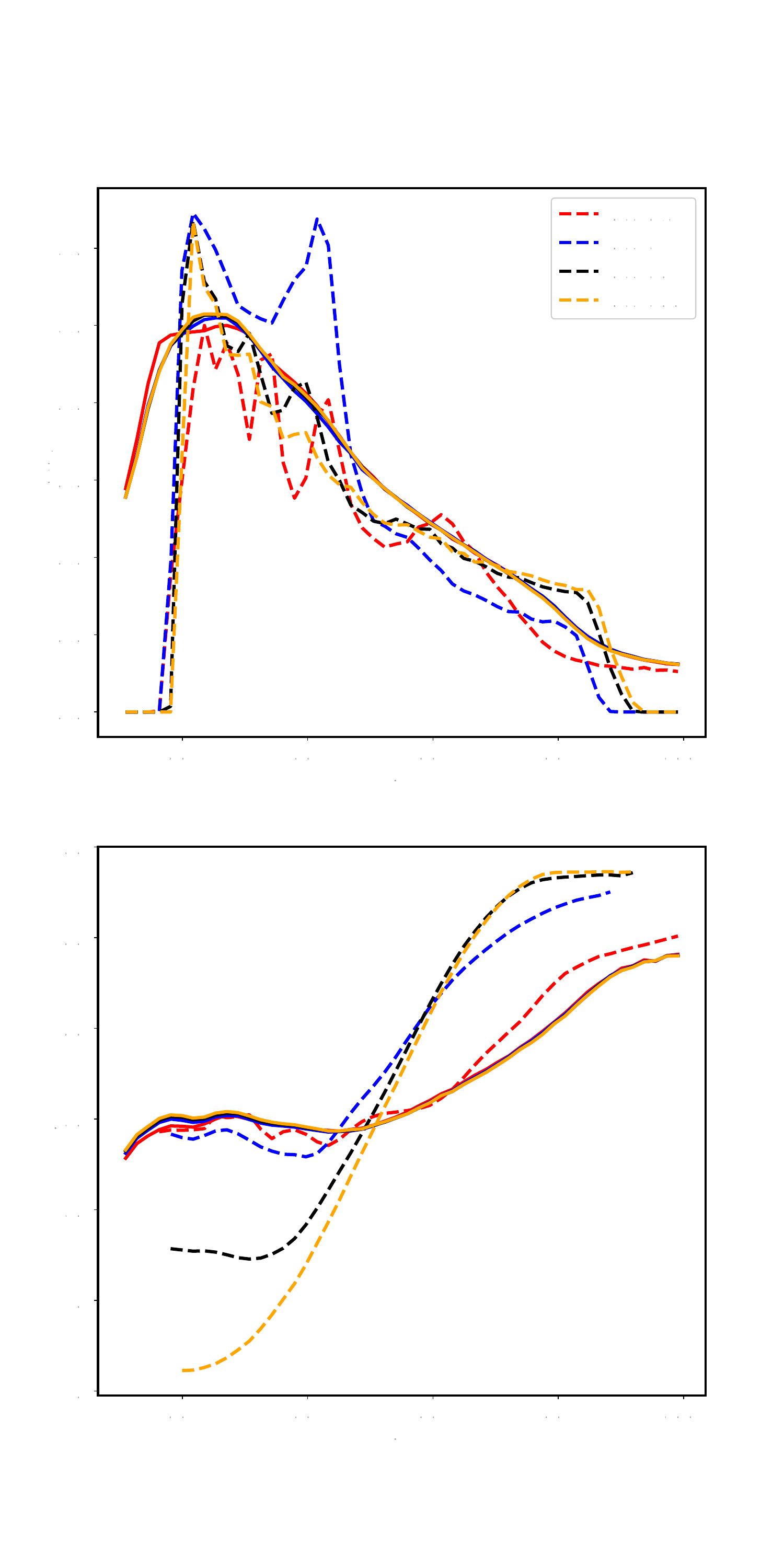}}}
 \caption{Top: shell averaged dust (dashed) and gas (solid) surface density radial profiles for the 4 snapshots shown in Figure~\ref{fig:splash_st}. Bottom: Inclination radial profiles for the same snapshots.\label{fig:profiles_st}
 }
 \end{center}
\end{figure}

Figure~\ref{fig:splash_st} shows the face-on column density of the dust component for the different dust sizes after $1$ orbit at the warp radius. We note the formation of density structures for all St values. The top panel of Figure~\ref{fig:profiles_st} shows the radial profiles of shell-averaged dust (dashed lines) and gas (solid lines) surface density of the $4$ different St simulations at the same time as Figure~\ref{fig:splash_st}. The oscillations in the azimuthally-averaged surface density quantitatively demonstrates the effects of WInDI. The gas surface density profiles are featureless, further demonstrating the lack of pressure bumps, and hence the absence of dust traps. We see the resulting density structures are stronger for St $\sim 0.1$ than the St > 1 simulations. The St $\sim 1$ simulation, as expected, suffers the most from the effects of radial drift as evident from the smaller dust disc extent. It also shows the narrowest density structures, which are not well resolved by our shell-averaging, complicating the quantitative comparison with the other St values. 

\begin{figure}
  \begin{center}
    \resizebox{78.0mm}{!}{\mbox{\includegraphics[angle=0]{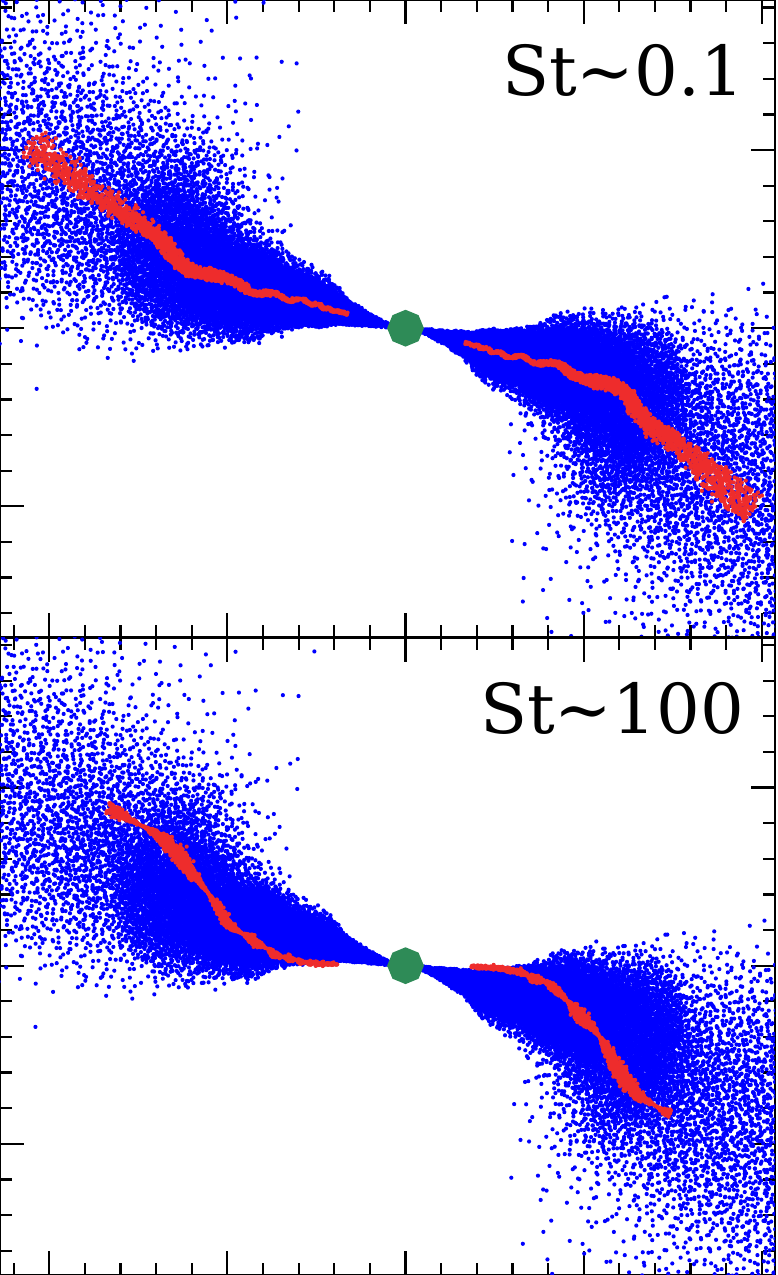}}}
 \caption{Cross section of the disc showing gas (blue) and dust (red) particles for the St $\sim 0.1$ (top) and St $\sim 100$ (bottom) cases\label{fig:particle_plot}.
 }
 \end{center}
\end{figure}

Figure~\ref{fig:particle_plot} shows the gas (blue) and dust (red) particle plots of a cross section in the disc at the same simulation time for the smallest (St $\sim 0.1$, top) and largest (St $\sim 100$, bottom) dust simulations. We see that small St dust closely follows the warp profile of the gas, while the warp in the large St dust only slowly evolves from the initial conditions (which is also clear from the bottom panel of Figure~\ref{fig:profiles_st} showing the tilt angle profiles). From this we deduce two different regimes for the evolution of dusty warps: an Aligned Dust regime at low St where the dusty warp closely follows the warp in the gas. In this regime, the role of the breathing and sloshing motions in forming dust structures is fairly straightforward and can be understood through the mechanism outlined in our toy model. Figure~\ref{fig:splash_time_evolution} and the leftmost panel in Figure~\ref{fig:splash_st} show that the dust structure in this regime is characterised by a break which has a constant orbital phase throughout the disc. This break occurs at the orbital phase where the sloshing displacement is minimum. The second regime is the Mis-Aligned Dust
where the St is high enough that the dusty warp does not follow the evolution of the gas warp, resulting
in a dust-gas misalignment. This regime is more complicated due to two reasons: 1) The gas sloshing (breathing) no longer acts on the dust horizontally (vertically). 2) the dust is thinner (see bottom panel of Figure~\ref{fig:particle_plot}), reducing the effect of the vertical dependence of
the sloshing/breathing oscillations. The dust still forms structure in this regime, but the interpretation
is not straightforward and does not result in the phase-coherent break which occurs in the Aligned Dust
regime (Figure~\ref{fig:splash_st}). The St at which the transition between the two regimes occurs will depend on how fast the gaseous warp is evolving.

\section{Discussion}
\label{section:discussion}
Our global SPH simulations show that WInDI is an effective way of globally accumulating dust in a warped disc. The resulting dust structures do not happen at gas pressure bumps, and therefore it is different from usual dust traps. We link this dust accumulation to the gas breathing and sloshing oscillations. In both our 1D and 3D investigations, WInDI appears to be triggered very quickly. However, our 3D simulations show, as expected, that the lifetime of dust structure resulting from WInDI is closely tied to the lifetime of the warp in the gas (note that this is is missing in our 1D analysis because the warp in this case is a coordinate transformation, and thus eternal). The relevance of WInDI in promoting dust growth and planetesimal formation is thus dependant on warp lifetime which is controlled by various disc parameters. Moreover, in this paper we only considered undriven warps imposed by our initial conditions. Driven warps, for example in a misaligned circumbinary disc, will have much longer warp lifetimes. We aim to further investigate this in a future study.

\subsection{In the context of other instabilities}
\citet{Ogilvie:2013lw} performed a linear stability analysis on the laminar solutions of the gas equations in the local warped shearing frame and concluded a parametric instability develops \citep{Paardekooper:2019ns,Fairburn:2023as}. In that work, they considered only axisymmetric perturbations, resulting in periodic time dependence of the linearised equations. We saw in Section~\ref{section:3D_results} that WInDI results in non-axisymmetric global structures, further complicating attempts to derive a growth rate from linear stability analysis. We thus leave this endeavour to future work and limit the scope of the current one to the reasoning based on our 1D toy model and the numerical results from our 3D SPH calculations. We note that our SPH simulations do not have enough resolution to capture the parametric instability \citep{Deng:2021fq}. It is likely that the onset of the parametric instability, which disrupts the regular laminar motions, might counteract the effects of WInDI. The resulting turbulent diffusion would also have complex effects that we aim to investigate in the future. On the other hand, it is possible that the dust back reaction in structures formed by WInDI might affect the onset of the parametric instability. This complex interplay between WInDI and the parametric instability warrants further detailed analysis. 

Our global simulations show that WInDI is effective at forming dust structures for St < 1 dust and small dust-to-gas ratio (0.01). This is a significant advantage over the Streaming Instability (SI) which requires larger dust particles and higher dust to gas ratios. Therefore, it is possible that WInDI may provide an easier route for planet formation when a warp is present, or indeed provide the conditions to trigger SI. We also note that, to date, the authors are not aware of any SPH study that was able to recover SI, possibly due to the method's inherent background viscosity inhibiting the onset of SI earliest stages, or perhaps the need for extremely high resolution. The fact that we had no problem recovering WInDI in our SPH simulations suggests, albeit tentatively, that WInDI might be more robust. Such an assertion needs more investigation and direct comparison between both instabilities, which we intend to perform in a future study.

Another relevant dust instability in proto-planetary discs is the Settling Instability. \citet{Squire:2018bb} first identified the Settling Instability as a member of the Resonant Drag Instabilities (RDI) framework (which also includes SI). The Settling Instability arises as a resonance between the dust vertical streaming as it settles towards the midplane, and the gas epicyclic frequency (while SI is due to the resonance between the gas epicyclic oscillations and dust radial and azimuthal streaming). It was found that the Settling Instability has a growth rate that is orders of magnitude larger than SI, does not need high dust-to-gas ratios, and is triggered for smaller dust grains as compared to SI. While the Settling Instability has not been widely reported in SPH studies, SPH practitioners often recover it in the initial phases of dusty discs simulations, for example in Figure 4 of \citet{Aly:2021bu} (blue curves indicating the dust surface density profiles for the earliest snapshot). The fact that  the Settling Instability is recoverable by SPH, but not SI, supports the finding that it has higher growth rates.

\begin{figure*}
  \begin{center}
    %\resizebox{78.0mm}{!}{\mbox{\includegraphics[angle=0]{Figures/WindiVsSettle.pdf}}}
    \includegraphics[width=0.8\textwidth]{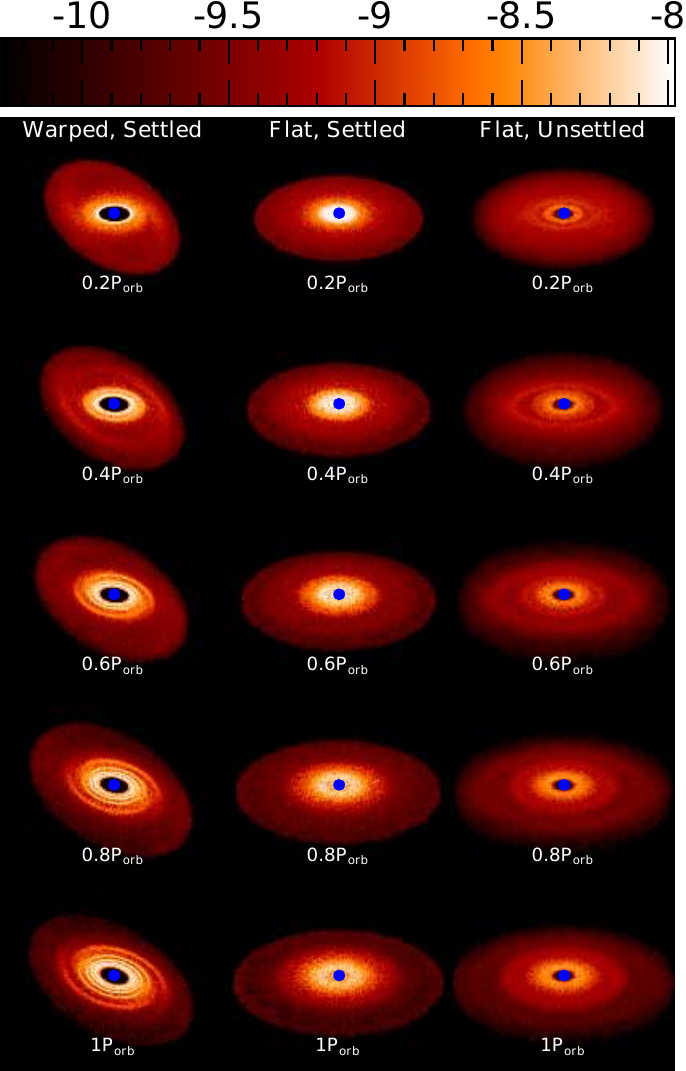}
 \caption{Dust (St=0.1) column density snapshots at various times (evolving top to bottom) of a warped disc (left column), a flat disc with the dust initially settled into the mid-plane, as with all other simulations in this paper so far (middle column), and a flat disc with the dust initially unsettled (right column). The warped case shows WInDI, the flat unsettled case (right) shows the settling instability, and the flat settled case (middle) shows neither. This shows that WInDI is distinct from the settling instability\label{fig:WindiVsSettle}
 }
 \end{center}
\end{figure*}

We perform an additional check to confirm that the dust structures we recover in our SPH simulations are indeed due to WInDI, not the Settling Instability. We recall that in our SPH setup we took extra care to minimise the effects of dust settling by starting the simulations with a dust thickness five times smaller than that of the gas. In Figure~\ref{fig:WindiVsSettle} we compare the dust column density for the St $\sim 0.1$ case (left column, `Warped, Settled') in our fiducial runs (same simulation as in Figure~\ref{fig:splash_time_evolution}) with two other reference simulations: a flat disc simulation with dust initially having a thickness five times smaller than the gas thickness (middle column, `Flat, Settled'), and another flat disc with dust thickness same as the gas (right column, `Flat, Unsettled'). The three simulations are identical in all other parameters. Figure~\ref{fig:WindiVsSettle} shows that the right column (Flat, Unsettled) recovers the Settling Instability as it forms dust rings. The middle column (Flat, Settled) shows minimal signs of these dust rings, if at all, which indicates that the smaller initial dust thickness is effective at reducing the Settling Instability. The left column (Warped, Settled) shows the initial phases of WInDI, which looks significantly different from the Settling Instability in the right column. This gives us confidence that WInDI is indeed distinct from the Settling Instability, even though both are easily recoverable with SPH.

\subsection{Caveats}
\label{section:limitations}
Our 1D toy model gives us insight into the the local effects that give rise to the cumulative compression cycles that lead to the instigation of WInDI. However, growth rates derived from this model are likely overestimated, since the imposed warp is a coordinate transformation and hence neither propagates nor dissipates. On the other hand, our global SPH 3D simulations likely underestimates the growth rate, owing to SPH deficiencies in capturing fluid instabilities due to the inherent particle reordering which results in an increased background effective viscosity. In future studies we aim to develop methods to derive more faithful local and global growth rates, for example using 3D local warped shearing box methods \citep{Paardekooper:2019ns} or 3D global moving mesh codes optimised to accurately model warps.

Another limitation to our analysis here is that the setup of our 3D SPH simulations starts with an idealised warp in the initial conditions, rather than create the warp during the simulation. Although this approach is not realistic, it helps us simplify the problem and isolate the effects of the warp on dust evolution. The significance of WInDI on dust evolution in real astrophysical systems can be better assessed in simulations where the warp is created throughout the simulation, for example by a flyby or later infall of misaligned material, or around misaligned circumbinary discs. We aim to investigate this in future studies.

\section{Conclusions}
\label{section:conclusion}
In this paper, We extend the warped shearing box framework developed by \citet{Ogilvie:2013by} to include dust and run 1D (vertical) calculations based on this framework. The local 1D analysis show that the gas `breathing mode' induced by the warp triggers a fast-growing dust instability, which is enhanced by the warp magnitude and dust Stokes number. The new instability (WInDI) is reproduced in our global 3D SPH simulations and results in dust substructures of enhanced density. We perform control simulations to show that WInDI is different from the dust Settling Instability but grows on a similar timescale. Our SPH simulations suggest that WInDI can globally manifest in two regimes: an Aligned Dust regime for small St where the warp in the dust follows that of the gas. In this regime the role of the breathing and sloshing modes is straightforward and the resulting substructure is characterised by broken rings. And a Mis-Aligned Dust regime for large St where the warp in the dust deviates from the gaseous warp. This regime is more complicated and the resulting dust substructure does not demonstrate breaks in the rings.

\section*{Acknowledgements}
The authors warmly thank Daniel J. Price for insightful comments and suggestions that improved the manuscript and Timothee David-Cléris for thorough discussions. We also thank the anonymous referee for comments that improved the clarity of the manuscript, in particular Figure 2. H.A. acknowledges funding from the European Research Council (ERC) under the European Union’s Horizon 2020 research and innovation programme (grant agreement No 101054502). R.N. acknowledges support from UKRI/EPSRC through a Stephen Hawking Fellowship (EP/T017287/1). J.-F.G.~acknowledges funding from the European Union's Horizon 2020 research and innovation programme under the Marie Sk\l{}odowska-Curie grant agreement No 823823 (DUSTBUSTERS) and from the ANR (Agence Nationale de la Recherche) of France under contract number ANR-16-CE31-0013 (Planet-Forming-Disks), and thanks the LABEX Lyon Institute of Origins (ANR-10-LABX-0066) for its financial support within the Plan France 2030 of the French government operated by the ANR. Rendered figures were made using \textsc{splash} \citep{Price:2007kx}. This work was performed on the OzSTAR national facility at Swinburne University of Technology. The OzSTAR program receives funding in part from the Astronomy National Collaborative Research Infrastructure Strategy (NCRIS) allocation provided by the Australian Government, and from the Victorian Higher Education State Investment Fund (VHESIF) provided by the Victorian Government. 

%%%%%%%%%%%%%%%%%%%%%%%%%%%%%%%%%%%%%%%%%%%%%%%%%%
\section*{Data Availability}
The data underlying this article will be shared on reasonable request to the corresponding author. The code \textsc{Phantom} used in this work is publicly available at https://github.com/danieljprice/phantom.



\bibliographystyle{mnras}
\bibliography{RN} % if your bibtex file is called example.bib

%\begin{thebibliography}{99}
%\end{thebibliography}

%%%%%%%%%%%%%%%%%%%%%%%%%%%%%%%%%%%%%%%%%%%%%%%%%%

%%%%%%%%%%%%%%%%% APPENDICES %%%%%%%%%%%%%%%%%%%%%

\appendix

\section{Dust density growth rates in a non-warped disc}
\label{section:dustgrowthflatdisc}
\begin{figure*}
    \centering
    \includegraphics[width=\textwidth]{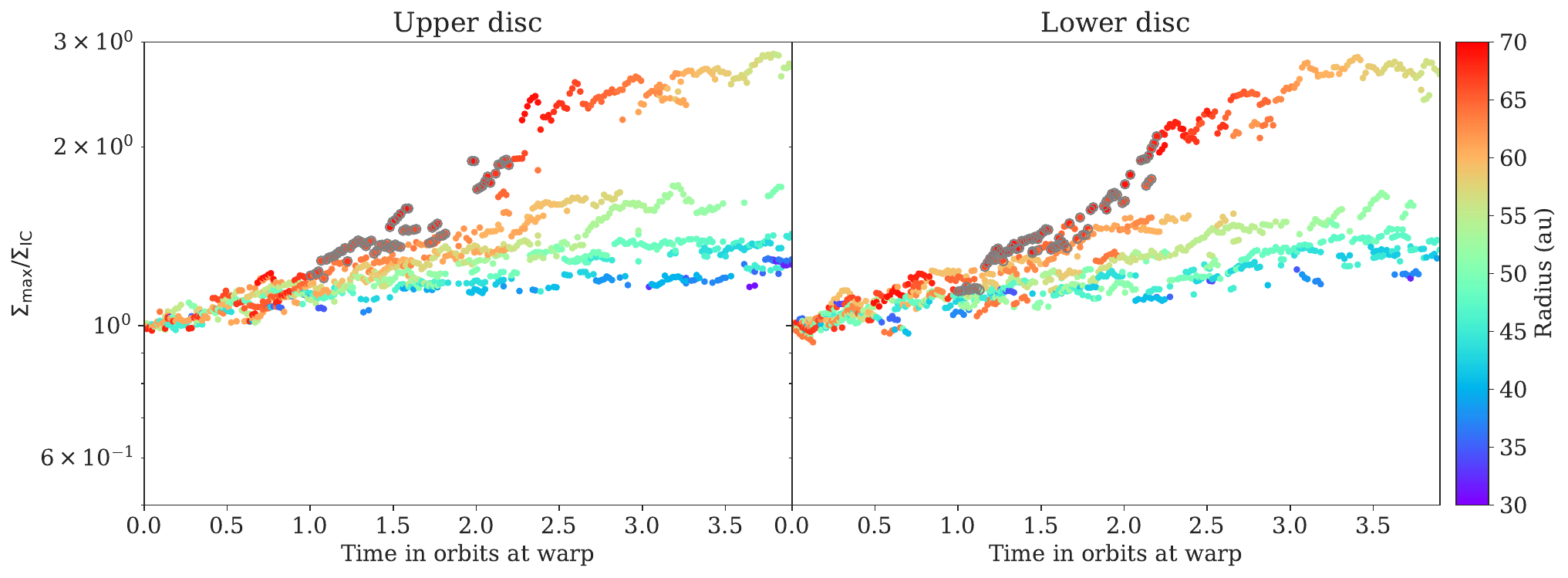}
    \caption{Growth of local dust concentrations in a flat, non-warped disc. Here we see steady growth at all radii (slightly faster at larger radii) in the dust surface density profile suggestive of radial drift.}
    \label{fig:flat_growth_rates}
\end{figure*}

Section~\ref{section:measuring_dust_growth} describes our process for measuring local dust enhancements and the growth of these throughout our simulation. In Figure~\ref{fig:growth_rates} we showed the growth rate as measured for our warped disc. In Figure~\ref{fig:flat_growth_rates} we use the same method in a disc that is identical but with no warp - that is, a dusty flat disc.

We identify two points of distinction between the dust density growth in our warped vs. flat disc. First, growth in the flat disc begins immediately whereas it is delayed slightly in the warped disc due to initial transients from the addition of the warp. Second, in the flat disc there is roughly constant growth at many radii simultaneously (at larger radii we do find slightly faster growth). By comparison, in the case of WInDI we see dust enhancements that start at small radii that then move outwards in radius as their concentration increases - i.e. a growing dust ring moving outwards. The growth we see here does not correlate to any global disc structures but is instead emblematic of global radial drift.

We calculate the growth rate in the flat disc case to demonstrate that the growth we see here is additionally slower than what we see in the warped case. We use points over the same time frame as in Figure~\ref{fig:growth_rates} and restrict ourselves to points that have a radius of more than $R=65$au, corresponding to the fastest growth in the flat disc case. These points are again indicated in Figure~\ref{fig:flat_growth_rates} with grey circles. From these we find an average growth rate of $0.554 \pm 0.001$/orbit. This is almost half the growth rate we see in the warped disc case and as mentioned above, is clearly distinguished from WInDI.

%%%%%%%%%%%%%%%%%%%%%%%%%%%%%%%%%%%%%%%%%%%%%%%%%%

% Don't change these lines
\bsp	% typesetting comment
\label{lastpage}
\end{document}